\documentclass[%
aps,prb,twocolumn,10pt,superscriptaddress, twocolumn,
showpacs,
 amsmath,amssymb,
 aps,
prb,
]{revtex4-1}
\usepackage{bpchem}
\usepackage{graphicx}
\usepackage{color}
\usepackage{multirow}

\begin{document}

\preprint{APS/123-QED}

\title{Moment evolution across the ferromagnetic phase transition of giant\\ magnetocaloric \BPChem{(Mn,Fe)\_{2}(P,Si,B)} compounds}% Force line breaks with \\

\author{H. Yibole}
 \affiliation{FAME, TUDelft, Mekelweg 15, 2629JB Delft, The Netherlands}
 \email{Yibole@tudelft.nl}
\author{F. Guillou}
\affiliation{FAME, TUDelft, Mekelweg 15, 2629JB Delft, The Netherlands}
\author{L. Caron}
 \affiliation{FAME, TUDelft, Mekelweg 15, 2629JB Delft, The Netherlands}
\author{E. Jim\'{e}nez}%
\affiliation{European Synchrotron Radiation Facility, 71 Avenue des Martyrs CS40220, F-38043 Grenoble Cedex 09, France}
\author{F.~M.~F.~de~Groot}%
 \affiliation{Inorganic Chemistry and Catalysis, Utrecht University, 3584 CG Utrecht, The Netherlands}
\author{P. Roy}
\affiliation{Electronic structure of Materials, Faculty of Sciences, Radboud University , 6525 AJ Nijmegen, The Netherlands}
\author{R. de Groot}
\affiliation{Electronic structure of Materials, Faculty of Sciences, Radboud University , 6525 AJ Nijmegen, The Netherlands}
\author{N.H. van Dijk}
\affiliation{FAME, TUDelft, Mekelweg 15, 2629JB Delft, The Netherlands}
\author{E. Br\"{u}ck }
 \affiliation{FAME, TUDelft, Mekelweg 15, 2629JB Delft, The Netherlands}

\date{\today}% It is always \today, today,
             %  but any date may be explicitly specified

\begin{abstract}
\ A strong electronic reconstruction resulting in a quenching of the Fe magnetic moments has recently been predicted to be at the origin of the giant magnetocaloric effect displayed by \BPChem{Fe\_{2}P}-based materials. To verify this scenario, X-ray Magnetic Circular Dichroism experiments have been carried out at the \textit{L} edges of Mn and Fe for two typical compositions of the \BPChem{(Mn,Fe)\_{2}(P,Si,B)} system. The dichroic absorption spectra of Mn and Fe have been measured element specific in the vicinity of the first-order ferromagnetic transition. The experimental spectra are compared with first-principle calculations and charge-transfer multiplet simulations in order to derive the magnetic moments. Even though signatures of a metamagnetic behaviour are observed either as a function of the temperature or the magnetic field, the similarity of the Mn and Fe moment evolution suggests that the quenching of the Fe moment is weaker than previously predicted. 

\end{abstract}

\pacs{Valid PACS appear here}% PACS, the Physics and Astronomy
                             % Classification Scheme.
%\keywords{Suggested keywords}%Use showkeys class option if keyword
                              %display desired
\maketitle

%\tableofcontents

\section{\label{intro} Introduction}

In the field of functional magnetic materials, compounds exhibiting a first-order magnetic phase transition (FOMT) have recently received a large interest due to their potential applications. Systems that undergo a discontinuous phase transformation have especially been studied for their magnetocaloric effect (MCE). One of the most promising MCE application is magnetic cooling, which overturns the use of greenhouse or ozone-depleting refrigerant gases, while it potentially has a better energy efficiency than usual cooling methods.~\cite{TIS03, GSC05, SMI12} To observe a particularly large magnetocaloric effect, advantage should be taken from the latent heat displayed by the first-order magnetic transition  leading to a “giant” magnetocaloric effect (GMCE). In the last decade several families of MCE materials have been discovered: \BPChem{FeRh},~\cite{ANN96} \BPChem{Gd\_{5}Si\_{2}Ge\_{2}},~\cite{PEC97} \BPChem{Mn(As,Sb)},~\cite{WAD01} \BPChem{La(Fe,Si)\_{13}} and its hydrides,~\cite{Hu01,FUJ03} \BPChem{MnFe(P,X)} X = As, Ge, Si, B.~\cite{TEG02, TRU10, DUN11, YIB14, GUI14} The understanding of the origin of the FOMT in these various materials is still limited. Especially the mechanism at the basis of the interplay between the magnetic/electronic/structural degrees of freedom that leads to a simultaneous change of all these parameters at the FOMT requires more insight and is crucial for further development of advanced magnetocaloric materials. As pointed out by several studies, the latent heat is the key parameter driving the GMCE properties.~\cite{GUI14, PER01, ANN04, SAN12} As an example, it has recently been reported that the substitution of a small amount of P by B in \BPChem{MnFe(P,Si)} materials results in a decrease of the latent heat, resulting in better magnetocaloric properties at intermediate magnetic fields.~\cite{GUI14} 

Many efforts have recently been made to understand the FOMT in \BPChem{Fe\_{2}P}-based \BPChem{(Mn,Fe)\_{2}(P,Si,B)} alloys. In this hexagonal system (space group $P\bar{6}2m$), the Fe and Mn atoms preferentially occupy the tetragonal 3\textit{f} site and the pyramidal 3\textit{g} site, respectively.~\cite{BEC91, DUN12}  When both sites are occupied by iron, a clear distinction is found between the low-moment 3\textit{f} site and the high-moment 3\textit{g} site.~\cite{BEC91} The FOMT of these materials has been extensively characterized, in particular, the structural evolution across the FOMT is now well documented.~\cite{LIA05, LIU09, HOG11, DUN12, OU13A, YUE13} All these studies support the magneto-elastic nature of the first-order ferro-to-paramagnetic transition, which leads to a discontinuity in the ratio of the cell parameters (\textit{c}/\textit{a}), but keeps the hexagonal structure unmodified with a negligible volume change. Recently, it has been proposed that the key parameter at the origin of GMCE is a strong electronic reconstruction at the FOMT. Electronic structure calculation on \BPChem{MnFeP\_{0.5}Si\_{0.5}}~\cite{DUN11} predicts a change in hybridization between the 3\textit{f} site, occupied by iron, and the surrounding non-metallic elements, which is expected to result in a reduction of the Fe(3\textit{f}) local moment from  \BPChem{1.54~{$\mu$}\_{B}} in the ferromagnetic phase to a value of only  \BPChem{0.003~{$\mu$}\_{B}} in the paramagnetic phase, while the moments on the 3\textit{g} site (Mn) are almost unaffected.~\cite{DUN11}Even though \BPChem{Fe\_{2}P} presents a FOMT one order of magnitude weaker than \BPChem{MnFe(P,Si)} alloys, a relatively similar mechanism was proposed by Yamada and Terao in the parent material.~\cite{YAM02} Within a Ginzburg-Landau model, the loss of long-range magnetic order at the FOMT was ascribed to a cooperative effect between the 3\textit{f} and 3\textit{g} sites, resulting in a significant reduction of the local 3\textit{f} moments at \BPChem{\textit{T}\_{C}}. These predictions have recently been revisited to take into account the influence of substitutional elements on the non-metallic site of \BPChem{Fe\_{2}P}.~\cite{GER13}

The evolution of the magnetic moments in \BPChem{Mn\_{1.25}Fe\_{0.7}P\_{0.5}Si\_{0.5}}, observed by neutron diffraction, seem to support this Fe-quenching scenario.~\cite{DUN12} In particular, a reduction of the local 3\textit{f} magnetic moment in the ferromagnetic phase is observed when \BPChem{\textit{T}\_{C}} is approached, which is compatible with a loss of the local magnetic moment at the FOMT. One of the aims of the present study is to pursue this analysis towards the paramagnetic regime. To test the predicted disappearance of the Fe moment in the paramagnetic phase, X-ray magnetic circular dichroism (XMCD) as a function of temperature and magnetic field has been measured. This method allows one to probe the evolution of the element-specific local moment \emph{both} in the ferromagnetic and paramagnetic phases. This is in contrast to neutron diffraction where site-specific moments are probed. This study has been carried out on two prototypical \BPChem{MnFe(P,Si,B)} materials: i) \BPChem{MnFe\_{0.95}P\_{0.582}B\_{0.078}Si\_{0.34}} (Composition \textit{A}), which exhibits a good MCE performance, with a Mn:Fe ratio close to 1, where Fe fully occupies the 3\textit{f} site and Mn the 3\textit{g} site; ii) a Mn-rich material \BPChem{Mn\_{1.25}Fe\_{0.70}P\_{0.50}B\_{0.01}Si\_{0.49}} (Composition \textit{B}), comparable with the composition used for the previous studies of the \BPChem{MnFe(P,Si)} system.~\cite{DUN12B, DUN12, YIB14}

\section{\label{exp}Experimental Methods}

Polycrystalline samples of \BPChem{MnFe\_{0.95}P\_{0.582}B\_{0.078}Si\_{0.34}} and \BPChem{Mn\_{1.25}Fe\_{0.70}P\_{0.50}B\_{0.01}Si\_{0.49}} were prepared according to the same method used in previous studies.~\cite{GUI14} The as-synthesized samples were cycled 5 times across the FOMT prior to the measurements. Magnetization measurements were carried out in a magnetometer (Quantum Design MPMS 5XL) equipped with a superconducting quantum interference device (SQUID) and the reciprocating sample option (RSO).

The X-ray magnetic circular dichroism (XMCD) measurements were collected at the ID08 beamline of the European Synchrotron Radiation Facility (ESRF) in Grenoble, France. The measurements were taken by tuning the energy at both the Mn and Fe \BPChem{\textit{L}\_{2,3}} edges ($2p$ $\to$ $3d$ transition). The X-ray absorption spectra were recorded using the Total-Electron Yield (TEY) mode, and normalized to the intensity of the incident beam. The sample temperature was regulated in the temperature range from 230 to 330~K. The X-ray absorption (XAS) spectra correspond to the sum of positive (${\mu^{+}}$)  and negative (${\mu^{-}}$) absorption signals, while the XMCD spectrum is calculated from the difference between ${\mu^{+}}$ and ${\mu^{-}}$. The bulk samples (circular disks with a diameter of 10 mm and a thickness of 2~mm) were placed in an ultra-high vacuum (UHV) system equipped with a 5~T split coil superconducting magnet. The incident X-ray beam and magnetic field are parallel to each other and oriented perpendicular to the sample surface. In order to ensure the cleanliness of sample surfaces and remove surface oxidation, the pellets were scrapped \emph{in situ} with a diamond file in the preparation chamber before the measurements. In order to reduce the occurrence of systematic errors, all measurement were performed for two directions of the applied magnetic field, along and opposite to the incident X-ray beam.

In order to simulate the 2\textit{p} XAS and XMCD, the spectra were modelled using the ligand field multiplet (LFM) theory, where we used the CTM4XAS interface version of the programs.~\cite{GRO90, GRO05, STA10} This approach takes into account all the electronic Coulomb interactions as well as the spin-orbit coupling on any electronic open shell and treats the geometrical environment of the absorbing atom through the crystal-field potential. The spectrum is calculated as the sum of all possible transitions for an electron excited from the 2\textit{p} level to a 3\textit{d} level. The 2\textit{p}4\textit{s} transitions are omitted as they have only small intensity. In the simplest formulation, a pure \BPChem{3\textit{d}\^n} configuration is attributed to the 3\textit{d} transition ions in the ground state and transitions between ${{2p}^6{3d}^n}$ ground state and ${{2p}^5{3d}^{n+1}}$ final excited state are calculated. The inter-electronic repulsions are introduced through Slater-Condon integrals, $F_{dd}^2$, $F_{dd}^4$ and the 3\textit{d} spin-orbit coupling (${\xi}_{3d}$) for the initial state and $F_{dd}^2$, $F_{dd}^4$, $F_{pd}^2$, $G_{pd}^1$, $G_{pd}^3$, ${\xi}_{3d}$ and ${\xi}_{2p}$ for the final state. Atomic spin-orbit values and Slater-Condon integrals have been used, where the Slater-Condon integrals are calculated via a 80\% reduction of the Hartree-Fock values. The surrounding of the metal ion is represented by an octahedral crystal field potential, parameterized by the parameter $10Dq$ that defines the energy difference between the $t_{2g}$ and $e_g$ orbitals. A molecular field ($\mu_{B}H$) of 0.02~eV is added along the \textit{z}-direction to describe the magnetic order.

The DFT calculations were done by using WIEN2k~\cite{BLA01}, which employs the full potential linearized augmented plane wave (FP-LAPW) method~\cite{SIN94}. The calculations were performed within a scalar relativistic mode. The non-overlapping atomic sphere radii were taken as 2.23 a.u., 2.45 a.u., 2.08 a.u. and 1.98 a.u. for Fe, Mn, Si and P respectively. The Brillouin zone integration was performed with the tetrahedron method with 72 $k$ points within the Irreducible Brillouin Zone (IBZ). Exchange interactions were taken into account using generalized gradient approximation (GGA) by Perdew, Burke and Ernzerhof (PBE).~\cite{PER96} To model the XMCD spectra, we first converged the spin-polarized calculation. Since the calculation of optical properties require a dense mesh of eigenvalues and eigenvectors,~\cite{AMB06} we chose a 112~$k$-points mesh in the IBZ. Then the spin-orbit calculation is performed via a second variational scheme with the direction of magnetization specified along the crystallographic \textit{z}-axis. Core states were treated fully relativistically. So states with orbital angular momentum $l\neq{0}$, show splitting due to spin-orbit interaction. To obtain the XMCD peaks at the correct energies, we considered the Slater transition state. So, 1/2 electron was removed either from the $2P_{3/2}$ or {$2P_{1/2}$} state and added to the valence states. The energy and the charge were converged to 0.0005 Ry and 0.005 electrons, respectively. Then the momentum matrix elements between the specific core states and the conduction states were calculated and finally the integration over the Brillouin zone was done.

\section{\label{results}Results and discussion}

\begin{figure}
     \centering
     \includegraphics[trim = 15mm 38mm 10mm 30mm, clip,width=0.58\textwidth]{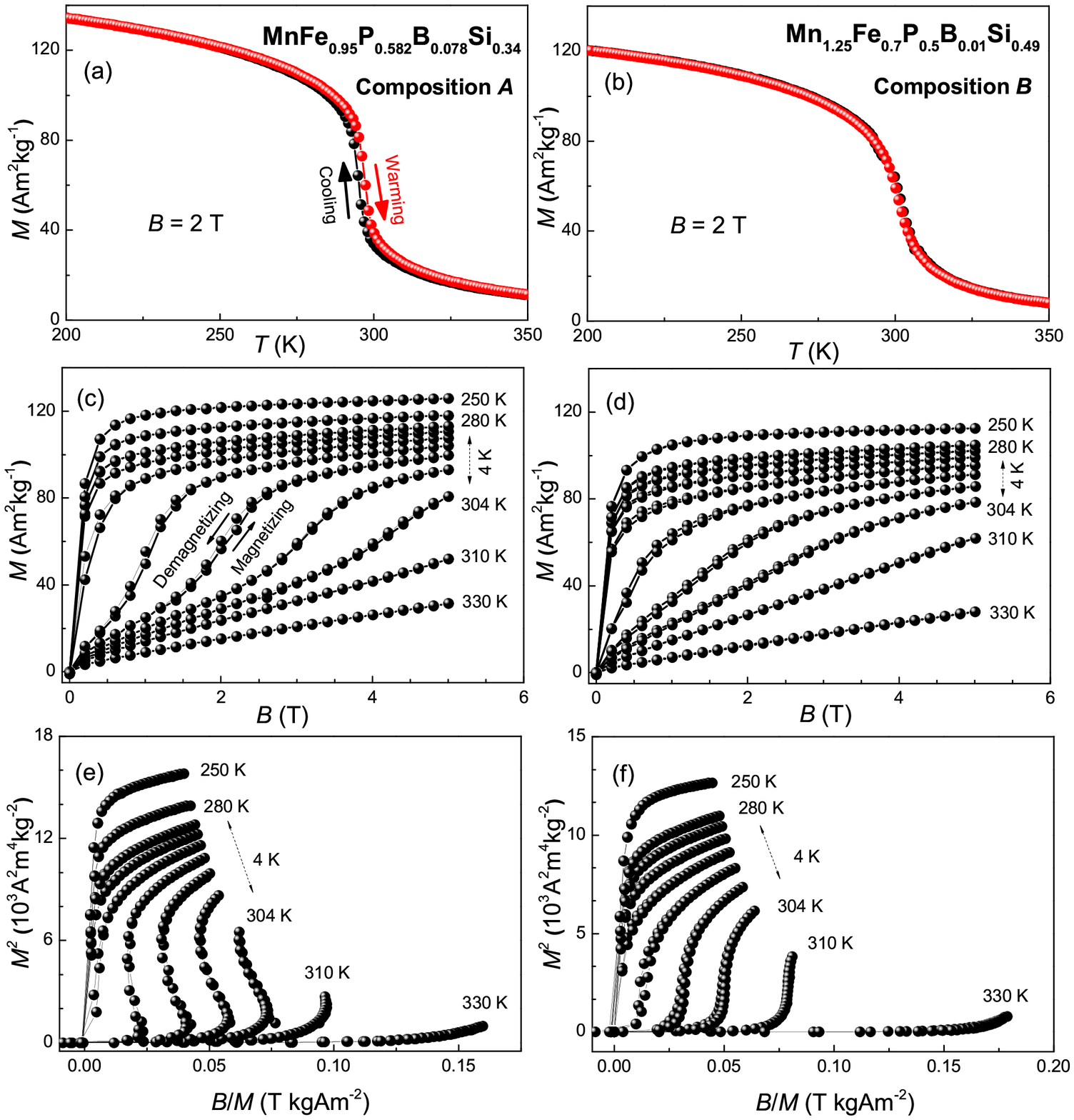}
     \caption{Temperature dependence of the magnetization in a field of 2 T for \BPChem{MnFe\_{0.95}P\_{0.582}B\_{0.078}Si\_{0.34}} (composition \textit{A}) (a) and \BPChem{Mn\_{1.25}Fe\_{0.70}P\_{0.50}B\_{0.01}Si\_{0.49}} (composition \textit{B}) (b), the Magnetic isotherms in the vicinity of the transition temperature for composition \textit{A} (c), and composition \textit{B} (d), and Arrot plots obtained from increasing field magnetization isotherms in the vicinity of the transition temperature for composition \textit{A} (e) and composition \textit{B} (f).}
     \label{fig1}
\end{figure}

The temperature dependence of the magnetization for \BPChem{MnFe\_{0.95}P\_{0.582}B\_{0.078}Si\_{0.34}} (Composition \textit{A}) and \BPChem{Mn\_{1.25}Fe\_{0.70}P\_{0.50}B\_{0.01}Si\_{0.49}} (Composition \textit{B}) are plotted in Figure \ref{fig1}(a) and \ref{fig1}(b). In agreement with previous reports, ~\cite{GUI14, DUN11} the Curie temperatures derived from \BPChem{\textit{M}\_{B}(\textit{T})} magnetization measurements in 2~T, are \BPChem{\textit{T}\_{C}} = 295 and 298~K for compositions \textit{A} and \textit{B}, respectively. A limited thermal hysteresis ($\sim2$~K) is noticeable between the \BPChem{\textit{M}\_{B}(\textit{T})} upon warming and upon cooling for composition \textit{A}. For composition \textit{B} the hysteresis is negligible. In addition, the magnetization jump appears to be broader in \textit{B} than in \textit{A}. Both features point towards a weaker first-order character in material \textit{B} than in \textit{A}. This tendency is supported by the isothermal magnetization curves presented in Figure \ref{fig1}(c) and \ref{fig1}(d) for materials \textit{A} and \textit{B}, respectively. The \BPChem{\textit{M}\_{T}(\textit{B})} curves for sample \textit{A} with Mn:Fe ratio close to 1 show a clear S-shape in the vicinity of \BPChem{\textit{T}\_{C}}, which confirms the occurrence of a FOMT. This metamagnetic behaviour is far less pronounced for material \textit{B}. The Arrot plots in Figure \ref{fig1}(e) and \ref{fig1}(f) indicate that sample \textit{B} lays at the boundary where the FOMT turns into a continuous transition. The two examples have similar Curie temperatures, but exhibit different transitional behaviors. 

\begin{figure}
     \centering
     \includegraphics[trim = 40mm 32mm 10mm 24mm, clip,width=0.53\textwidth]{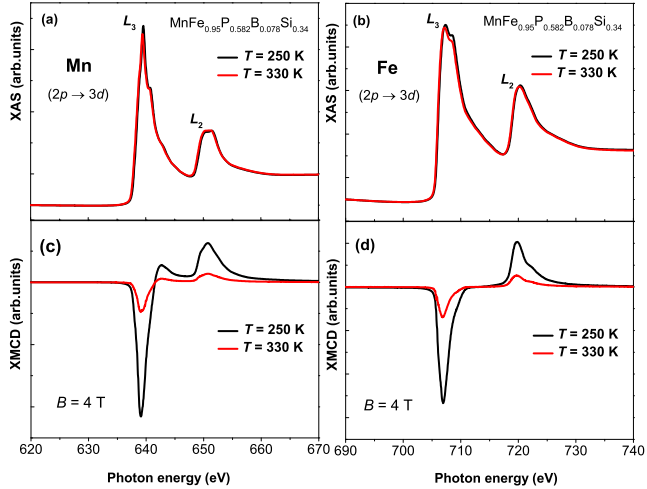}
     \caption{XAS and XMCD spectra for \BPChem{MnFe\_{0.95}P\_{0.582}B\_{0.078}Si\_{0.34}} (composition \textit{A}) measured at the Mn-\BPChem{\textit{L}\_{2,3}} edge (left panel) (a) and (c), and Fe-\BPChem{\textit{L}\_{2,3}} edge (right panel) (b) and (d). Black and red spectra are measured at 250~K (ferromagnetic state) and 330 K (paramagnetic state), under an applied magnetic field of 4~T, respectively.}
     \label{fig2}
\end{figure}

Figure \ref{fig2} shows the experimental XAS and XMCD spectra of composition \textit{A} at \BPChem{\textit{L}\_{2,3}} edges of Mn and Fe. The spectra were obtained in the ferromagnetic state (\textit{T} = 250~K) and in the paramagnetic state (\textit{T} = 330~K) at an applied magnetic field of 4~T. There is almost no difference between the XAS spectra of the ferromagnetic and paramagnetic state (see Figure \ref{fig2}(a)). For Mn, the absorption maxima are located at about 639.6 and 650.8~eV for the \BPChem{\textit{L}\_{3}} and \BPChem{\textit{L}\_{2}} edges, respectively. On the high energy wing of \BPChem{\textit{L}\_{3}}, one can observe two satellite peaks, one at about 1~eV and another (hardly visible) at about 3.5~eV above the maximum of \BPChem{\textit{L}\_{3}}. At the \BPChem{\textit{L}\_{2}} edges, there is a tendency towards a peak splitting into two separate peaks. The two peaks are also present in the multiplet calculation of Mn, regarded as a $4s^{2}3d^{5}$ system. Due to band dispersion effects these atomic transitions are broadened in intermetallic and covalent systems. Depending on the degree of broadening the peaks can be distinguishable in experimental spectra, where Mn is more localized and expected to have more visible features.~\cite{GRO05, EDM06} Similar features were also observed for the Mn spectra of some intermetallic alloys and thin films.~\cite{KLA11, PRI13, DUR97} This more localized character of the 3\textit{g} site in \BPChem{Fe\_{2}P}-based materials has already been pointed out by several theoretical and experimental studies.~\cite{DUN11, GER13, CAR14} However, at this stage one cannot rule out a few other possibilities for the origin of the peak splitting. Although the sample surface was prepared \emph{in situ} in UHV, one has to keep in mind that a partial oxidation of the surface cannot be totally discarded, especially when measuring in TEY mode (the probing depth is only a few tens of a nm). In addition, there may be an antisite effect, \emph{i.e.} a limited amount of Mn site on the 3\textit{f} tetrahedral site instead of the 3\textit{g} site with a pyramidal environment.~\cite{DUN12} In contrast to the Mn case, the XAS line shapes of Fe, shown in Figure \ref{fig2}(b), exhibit a rather different behavior with relatively broad absorption peaks. This suggests that the Fe atoms are in a more delocalized environment than the Mn atoms in this system. The only additional feature is a small peak that appears on the high-energy site of the \BPChem{\textit{L}\_{3}} edge, positioned at about +1.1~eV. It is worth noting that for both the Fe and Mn XAS spectra, no energy shifts are observed when the temperature is changed from the ferromagnetic to the paramagnetic state. This implies that no significant valence change takes place across the phase transition.

Figure \ref{fig2}(c) and \ref{fig2}(d) show the XMCD spectra for sample \textit{A}. The intensity of both the Mn and Fe XMCD are clearly reduced in the paramagnetic state (330~K), while in both cases the spectral shape remains unmodified. Moreover, the polarity of the XMCD of Mn and Fe are the same, which indicates a parallel alignment of the spin moments of Mn and Fe. This is in good agreement with the neutron diffraction results for \BPChem{MnFe(P,Si)} compounds in which the spins on the 3\textit{f} and 3\textit{g} sites are aligned in parallel in the ab plane.~\cite{DUN12} For the Mn spectrum, there is a pronounced positive shoulder on the high-energy side of \BPChem{\textit{L}\_{3}} edge, whose intensity approaches zero before the \BPChem{\textit{L}\_{3}} peak. This positive shoulder at the vicinity of \BPChem{\textit{L}\_{3}} of the Mn spectrum was also found in atomic calculations for Mn $4s^{2}3d^{5}$states.~\cite{LAA91} The contribution at 3.5~eV above \BPChem{\textit{L}\_{3}}, which is visible in both the XMCD and the XAS signal, is considered to belong to the main phase and to be an intrinsic property of the material. This is a signature of the \textit{jj} mixing often observed in light 3\textit{d} elements and will be discussed in detail hereafter. Unlike the anomaly at \BPChem{\textit{L}\_{3}}+3.5~eV, the additional peaks at \BPChem{\textit{L}\_{3}}+1~eV in the XAS spectrum of both Fe and Mn, do not correspond to any feature in the XMCD signal. This suggests that oxidation is the origin of these two satellite peaks at +1 eV on both the Mn and Fe edges. These satellite peaks however do not show any variation with temperature or magnetic field in the XAS and XMCD signals, and will therefore not influence the discussion on the changes across the FOMT.
\begin{figure}
     \centering
     \includegraphics[trim = 20mm 30mm 20mm 22mm, clip,width=0.45\textwidth]{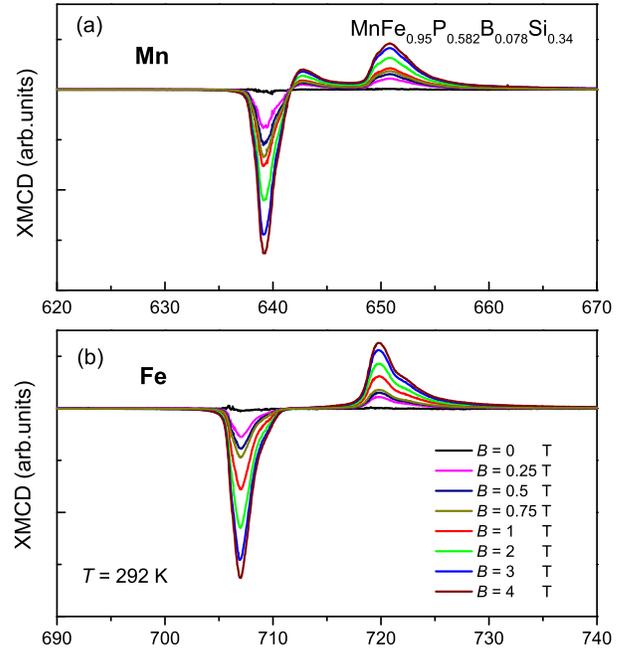}
     \caption{The field dependence of the XMCD spectra for Mn-\BPChem{\textit{L}\_{2,3}} (a) and Fe-\BPChem{\textit{L}\_{2,3}} (b) at a temperature of 292 K for \BPChem{MnFe\_{0.95}P\_{0.582}B\_{0.078}Si\_{0.34}} (composition \textit{A}). The temperature of 292~K is just above the magnetic transition at zero field of \BPChem{\textit{T}\_{C}}  = 288~K (\textit{B} = 0~T).}
     \label{fig3}
\end{figure}

Figure \ref{fig3}(a) and \ref{fig3}(b) show the field dependence of the XMCD spectra for Mn and Fe at a temperature of 292~K, just above the transition temperature \BPChem{\textit{T}\_{C}} = 288~K (in zero magnetic field). As anticipated, the XMCD intensity increases with the applied magnetic field, reflecting a gain in the average projection of the magnetic moment with the field. From \textit{B} = 0 to 0.75~T, there is a rapid appearance of a sizable XMCD signal, which can be ascribed to the orientation of magnetic domains. Above 0.75~T, one can see a slower increase of the XMCD signal with an approach to a saturation above 3~T. This behaviour corresponds to a metamagnetic transition. The critical field (\BPChem{\textit{B}\_{C}}) is reasonably in line with the \BPChem{\textit{B}\_{C}} $\approx$ 1.25~T derived from the \BPChem{\textit{M}\_{T}(\textit{B})} curves presented in Figure \ref{fig1}(c). Comparing Figure \ref{fig2} and \ref{fig3} highlights the similarity between crossing the paramagnetic-ferromagnetic transition by decreasing the temperature or by increasing the field in the present \BPChem{(\textit{T}, \textit{B})} range. For both the temperature- and field-induced FOMT, the shape of the XMCD (and XAS) spectra at the Mn and Fe edge is not affected by the change in magnetic field and temperature. 

\begin{figure}
     \centering
     \includegraphics[trim = 40mm 32mm 10mm 24mm, clip,width=0.53\textwidth]{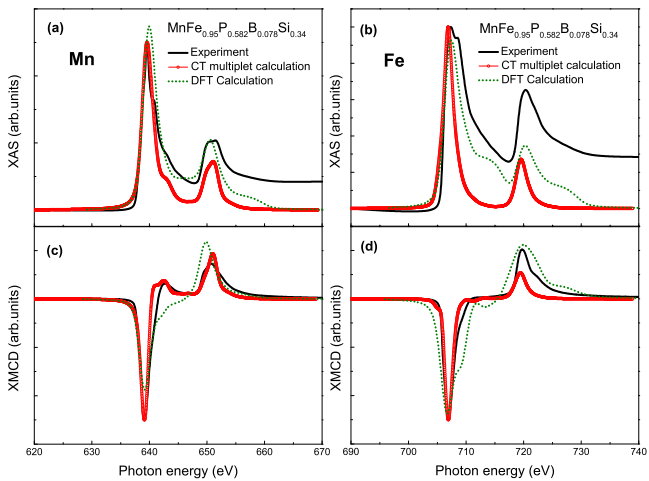}
     \caption{Comparison of the measured, Charge transfer multiplet calculation and DFT calculation for XAS and XMCD spectra at Mn-\BPChem{\textit{L}\_{2,3}} (a) and (c) and Fe-\BPChem{\textit{L}\_{2,3}} (b) and (d) edge, for \BPChem{MnFe\_{0.95}P\_{0.582}B\_{0.078}Si\_{0.34}} (composition \textit{A}). Black full line shows the measured spectra at 250~K, 4~T. Red symbols and green dots show the spectra from the multiplet and DFT calculations, as described in the text.}
     \label{fig4}
\end{figure}

The line shape of the XAS and XMCD spectra are strongly dependent on the electronic configuration of the probed atoms. In order to analyse the spectral features of the XAS and XMCD spectra in detail, Charge Transfer (CT) multiplet calculations~\cite{GRO90, GRO05, STA10} and Density Functional Theory (DFT) calculations were carried out. In Figure \ref{fig4}, we compare the experimental data at magnetic saturation for composition \textit{A} with the spectra obtained from the CT and DFT calculations. The DFT calculations can ideally provide a good single-particle (itinerant model) description of the chemical bonds, while the multiplet calculation provides a reliable multiconfigurational description of the final states and their spin-orbit coupling. In the present CT calculations, the spectra were modelled by the $2p^{6}3d^{5}$ $\to$ $2p^{5}3d^{6}$ (Mn) and $2p^{6}3d^{6}$ $\to$ $2p^{5}3d^{7}$ (Fe) transition in an octahedral \BPChem{\textit{O}\_h} symmetry, considering a crystal-field splitting of $10Dq$ = 0.2~eV for Mn and $10Dq$ = 1.0~eV for Fe. To account for configuration interaction effects, the Slater integrals were reduced to 80\% of their Hartree-Fock values. The effect of exchange splitting was taken into account by setting the magnetic splitting parameter to 20~meV. The final state charge transfer energy ${\Delta}$+\BPChem{\textit{U}\_{dd}}-\BPChem{\textit{U}\_{pd}} has been used the fixed difference of \BPChem{\textit{U}\_{pd}}-\BPChem{\textit{U}\_{dd}} = 1~eV, where ${\Delta}$ is the charge transfer energy, \BPChem{\textit{U}\_{dd}} is the Hubbard U correlation energy and \BPChem{\textit{U}\_{pd}} is the core-hole potential.~\cite{ZAA85} Each spectrum is broadened with a Lorentzian broadening of 0.2~eV (0.4~eV) for \BPChem{\textit{L}\_{3}}(\BPChem{\textit{L}\_{2}}) and a Gaussian broadening of 0.5~eV to approximately account for lifetime and resolution effects. The XMCD was derived from DFT calculations according to the method described in section \ref{exp}.

These two computational methods are compared to the experimental XAS and XMCD data in Figure \ref{fig4}(a)/\ref{fig4}(c) and \ref{fig4}(b)/\ref{fig4}(d) for the Mn and Fe edges, respectively. Generally speaking, there is a decent agreement between the experimental spectra and the calculations. One can see that the overall spectral features  are reproduced fairly well by both calculation methods. The energy positions of the main peaks of the two calculated spectra agree with that of experimental spectra. Besides, the main and satellite features are reproduced correctly within 1~eV and the calculated intensity distributions are rather realistic (see Figure \ref{fig4}(a) and \ref{fig4}(b)). However, one observes that the relative intensity ratio \BPChem{\textit{L}\_{2}}/\BPChem{\textit{L}\_{3}}, which depends mainly on the projection of the total moment, shows some deviation for both calculations (see Figure \ref{fig4}(c) and \ref{fig4}(d)). Furthermore, looking more closely into the three spectra, the multiplet calculation seems to yield a better overlap with the experiment for Mn than the DFT method. In particular, the multiplet calculation succeeds in reproducing the positive shoulder on the high-energy side of \BPChem{\textit{L}\_{3}} edge, while the DFT fails in doing so. In contrast, for Fe, there is a better agreement between the spectrum calculated from DFT and the experiment. Especially, the ratio of the \BPChem{\textit{L}\_{2}}/\BPChem{\textit{L}\_{3}}, XMCD peaks are more consistent with DFT than CT calculations. In addition, for Fe, the width of the peaks from multiplet calculation is too narrow in comparison with experiments and DFT calculation. This reduction in overall width of the peaks in multiplet calculation can be taken into account by a reduction of the Slater integrals.~\cite{LAA86} The differences between Mn and Fe spectra suggest that although Mn and Fe atoms are closely related $3d$ transition metals, they behave intrinsically differently in this system. More precisely, it is anticipated from previous studies that the Mn $3d$ states occupy a more localized environment (the pyramidal $3g$ site), while Fe $3d$ states are in a more delocalized one (tetrahedron $3f$ site).~\cite{CAI93} This may explain why the charge transfer multiplet calculation cannot fully be implemented in the case of Fe.

To derive quantitative values for the spin and orbital moments (\BPChem{{$\mu$}\_{spin}} and \BPChem{{$\mu$}\_{orb}}) from XMCD spectra, sum rules are usually applied.~\cite{THO92, CAR93} However, in the present case, some care has to be taken, especially for Mn.~\cite{PIA09, YOS96, LAA04, WU93, CRO96} Several issues neglected by the sum rule approach play a significant role. First of all, the relatively strong $2p$-$3d$ (core-valence) Coulomb interaction compared to $2p$ spin-orbit interaction leads to mixing of the \textit{j} = $2p_{3/2}$ and \textit{j} = $2p_{1/2}$ manifolds, which consequently causes an inaccuracy for the integration over the spin-orbit split core levels. This feature is illustrated by the positive XMCD signal on the high energy wing of \BPChem{\textit{L}\_{3}}. The spin sum rule is only valid in the limit of \BPChem{\textit{jj}} coupling and in the present case should be thus corrected for \BPChem{\textit{jj}} mixing. Second, according to the sum rules, \BPChem{{$\mu$}\_{spin}} and \BPChem{{$\mu$}\_{orb}} depend on the number of valence holes in the $3d$ state and a proportionality constant, the integrated area of magnetization-averaged signal. The former can be determined via theoretical calculations, but the latter requires care to obtain background subtraction and accurate edge steps, which often causes inaccuracies. Finally, the sum rules are often applied by omitting the magnetic dipole operator \BPChem{\textit{T}\_{z}}. For the present case, this assumption is applicable, as for $3d$ metals it is proved to be negligible.~\cite{PIA09, CHEN95, WU93} Here, we assumed the number of unoccupied $d$ holes to be 5 for Mn and 4 for Fe. Following the procedure used originally by Chen \emph{et al.},~\cite{CHEN95} we employed a simple two-step function to subtract the \BPChem{\textit{L}\_{3}} and \BPChem{\textit{L}\_{2}} edge steps from the absorption spectrum. Then by applying orbital sum rules,~\cite{THO92} the orbital magnetic moments of Mn and Fe of composition \textit{A} and \textit{B} are obtained and presented in Table \ref{table1}. A small positive orbital moment is observed for both Mn and Fe, which confirms that the spin and orbital moments are coupled in parallel. Besides, the small nonzero value also indicates that Mn and Fe are not in a pure $3d^5$ and $3d^6$ ground state configuration, but have a small admixtures of $3d^6$ for Mn and $3d^7$ for Fe due to the hybridization with neighbouring atoms.~\cite{LAA10, KRO06} Though both the Mn and Fe atoms have a very small orbital magnetic moment, it is interesting to note that \BPChem{{$\mu$}\_{orb}} for Fe is one order of magnitude higher than for Mn. 

To obtain more reliable spin moments, a correction procedure is used.~\cite{PIA09} First, an experimental spin moment is derived from the sum rules.~\cite{CAR93} Then, XAS and XMCD are simulated with charge-transfer multiplet calculations by fitting the experimental data. At the end, the expected spin moments are calculated for the ground state. By comparing this value to the sum rules result, correction factors~\cite{PIA09} [$SE_{z}^{sum}$]/$\langle$$S_{z}$$\rangle$ of 0.69 for Mn and 0.85 for Fe are derived, and are subsequently applied to all sum-rule values. In Table \ref{table1}, the spin and orbital magnetic moments derived from the sum rules and the corrected values are summarized along with the total magnetic moment obtained by SQUID magnetometry. For composition \textit{A}, we obtain an effective magnetic moment for Mn of $\mu$(Mn) = \BPChem{{$\mu$}\_{spin}}+ \BPChem{{$\mu$}\_{orb}}= \BPChem{1.185~{$\mu$}\_{B}} and for Fe atom of $\mu$(Fe) = \BPChem{{$\mu$}\_{spin}}+ \BPChem{{$\mu$}\_{orb}} = \BPChem{0.838~{$\mu$}\_{B}} in the ferromagnetic state (\textit{T} = 250~K, 2~T). The total magnetic moment per formula unit results in \BPChem{2.023~{$\mu$}\_{B}}, which is about 30\% smaller than the magnetization determined by magnetometry of \BPChem{2.96 {$\mu$}\_{B}/f.u.}. In the paramagnetic state (\textit{T} = 330~K, 2~T),  Mn moments of $\mu$(Mn)  = \BPChem{{$\mu$}\_{spin}}+ \BPChem{{$\mu$}\_{orb}} = \BPChem{0.311~{$\mu$}\_{B}} and Fe moments of $\mu$(Fe) = \BPChem{{$\mu$}\_{spin}}+ \BPChem{{$\mu$}\_{orb}}= \BPChem{0.235~{$\mu$}\_{B}} are obtained. For composition \textit{B}, a slightly lower magnetic moment is observed, which is in accordance with the magnetization data shown in Figure \ref{fig1}.
 
\begin{table*}
\caption{\label{table1}The spin and orbital magnetic moments (in units of \BPChem{{$\mu$}\_{B}}/atom) derived from the sum rules and corrected spin-moment values are summarized along with the total magnetic moment obtained by SQUID magnetometry, for \BPChem{MnFe\_{0.95}P\_{0.582}B\_{0.078}Si\_{0.34}} (composition \textit{A}) and \BPChem{Mn\_{1.25}Fe\_{0.70}P\_{0.50}B\_{0.01}Si\_{0.49}} (composition \textit{B}) at 250~K (ferromagnetic state) and 330~K (paramagnetic state) in a field of 2~T.}
\begin{ruledtabular}
\begin{tabular}{cccccccccccc}
\multirow{3}{*}{Samples}                                                            & \multirow{3}{*}{Atom}        & \multicolumn{4}{c}{Ferromagnetic \textit{T} = 250 K}           & \multicolumn{4}{c}{Paramagnetic \textit{T} = 330 K}            & \multicolumn{2}{c}{$\mu$(SQUID)}                  \\ \cline{3-12} 
                                                                                    &                              & \multicolumn{2}{c}{\BPChem{$\mu$\_{spin}}} & {\BPChem{$\mu$\_{orb}}}  & {\BPChem{$\mu$\_{orb}}}/{\BPChem{$\mu$\_{spin}}}       & \multicolumn{2}{c}{\BPChem{$\mu$\_{spin}}} & {\BPChem{$\mu$\_{orb}}}  & {\BPChem{$\mu$\_{orb}}}/{\BPChem{$\mu$\_{spin}}}       & \multirow{2}{*}{250 K} & \multirow{2}{*}{330 K} \\
                                                                                    &                              & {[}SR{]}    & {[}Corr.{]}    & {[}SR{]} & {[}Corr.{]} & {[}SR{]}    & {[}Corr.{]}    & {[}SR{]} & {[}Corr.{]} &                        &                        \\ \hline
\multirow{2}{*}{\BPChem{MnFe\_{0.95}P\_{0.582}B\_{0.078}Si\_{0.34}}}       & $\langle$ Fe $\rangle$ \textit{3f}    & 0.671       & 0.792          & 0.046    & 0.058       & 0.190       & 0.223          & 0.012    & 0.054       & \multirow{2}{*}{2.96}  & \multirow{2}{*}{0.37}  \\
                                                                                    & $\langle$ Mn $\rangle$ \textit{3g}   & 0.816       & 1.184          & 0.001    & 0.001       & 0.214       & 0.310          & 0.001    & 0.003       &                        &                        \\ \hline
\multirow{2}{*}{\BPChem{Mn\_{1.25}Fe\_{0.7}P\_{0.5}B\_{0.01}Si\_{0.49}}} & $\langle$ Fe $\rangle$ \textit{3f}    & 0.658       & 0.776          & 0.023    & 0.030       & 0.120       & 0.174          & 0.015    & 0.086       & \multirow{2}{*}{2.68}  & \multirow{2}{*}{0.31}  \\
                                                                                    & $\langle$ Mn $\rangle$ \textit{3f}+\textit{3g} & 0.739       & 1.072          & 0.006    & 0.006       & 0.172       & 0.250          & 0.005    & 0.020       &                        &                        \\ 
\end{tabular}
\end{ruledtabular}
\end{table*}

\begin{figure}
     \centering
     \includegraphics[trim = 5mm 20mm 20mm 15mm, clip,width=0.45\textwidth]{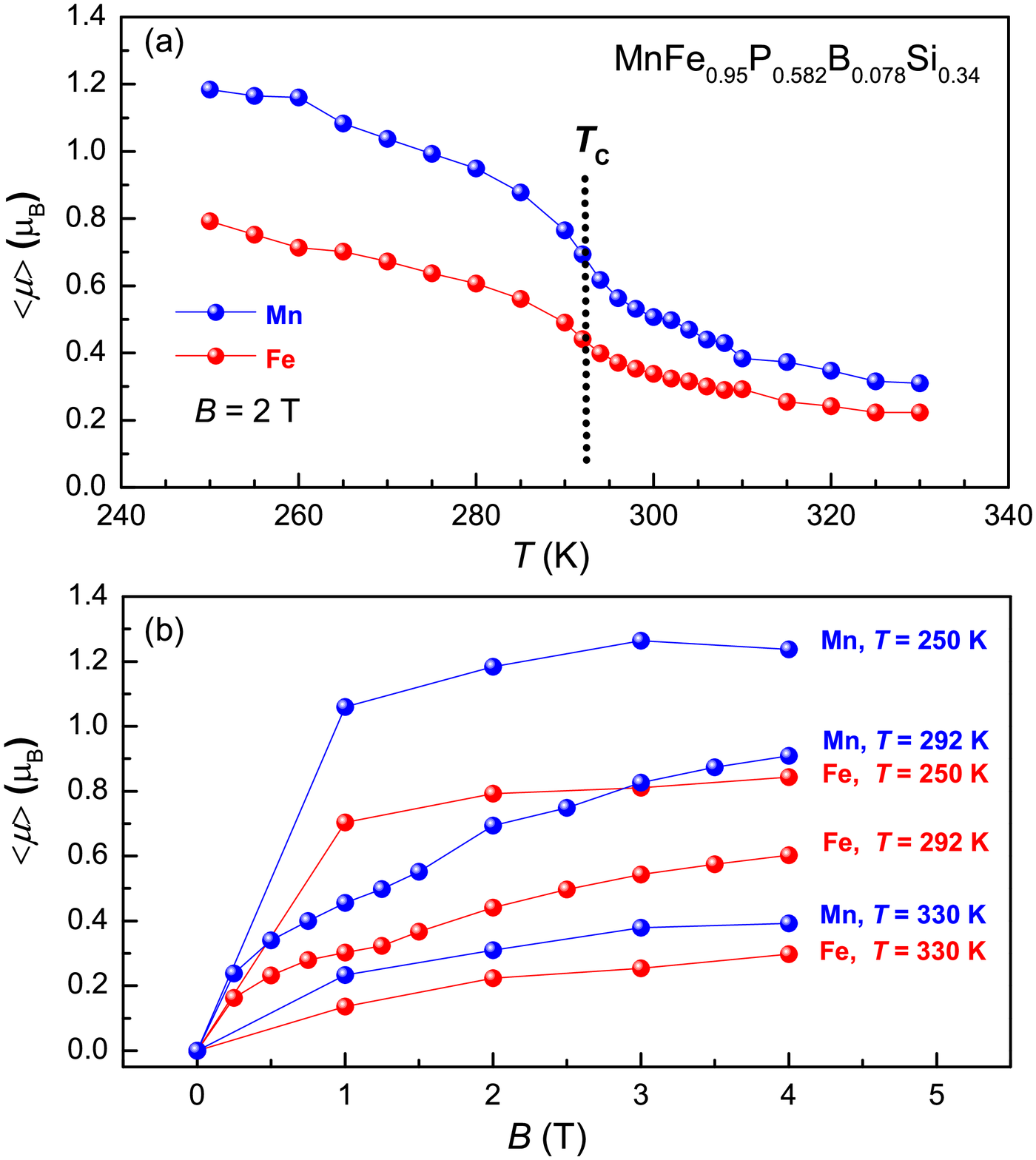}
     \caption{The XMCD magnetic moment as a function of temperature (a) and magnetic field (b) for \BPChem{MnFe\_{0.95}P\_{0.582}B\_{0.078}Si\_{0.34}} (composition \textit{A}). Top curves were measured in a field of 2~T, below curves were measured at 250~K (Ferromagnetic state), 292~K (near transition) and 330~K (Paramagnetic state). The XMCD magnetic moments were derived as described in the text.}
     \label{fig5}
\end{figure}

\begin{figure}
     \centering
     \includegraphics[trim = 5mm 15mm 20mm 20mm, clip,width=0.45\textwidth]{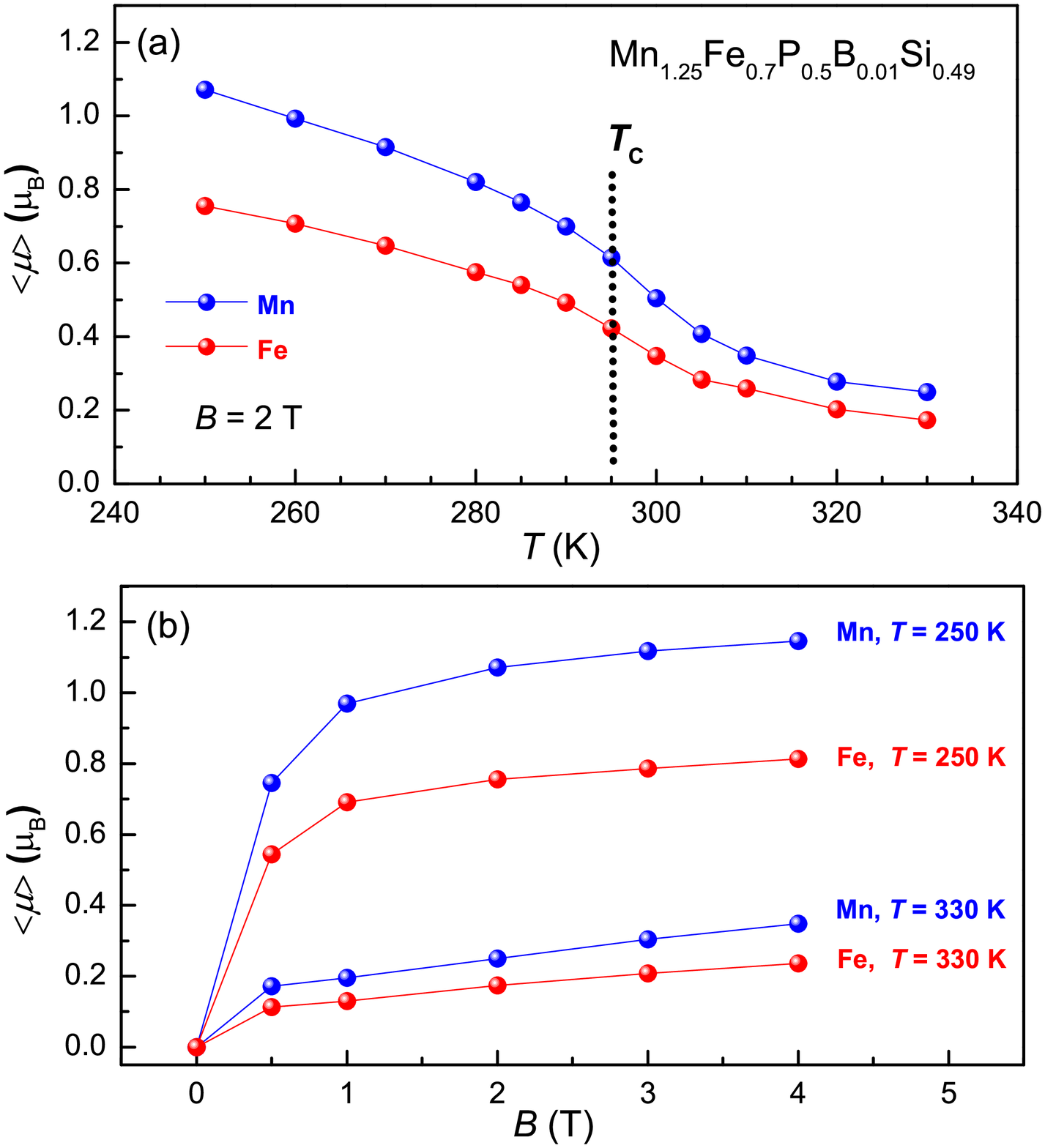}
     \caption{The XMCD magnetic moment as a function of temperature (a) and magnetic field (b) for \BPChem{Mn\_{1.25}Fe\_{0.70}P\_{0.50}B\_{0.01}Si\_{0.49}} (composition \textit{B}). Top curves were measured in a field of 2~T, below curves were measured at 250~K (Ferromagnetic state) and 330~K (Paramagnetic state). The XMCD magnetic moments were derived as described in the text.}
     \label{fig6}
\end{figure}

To gain further insight into the thermal evolution of the magnetic properties across the FOMT, systematic XMCD measurements were performed as a function of the temperature and magnetic field. The results are presented in Figures \ref{fig5} and \ref{fig6} for materials \textit{A} and \textit{B}, respectively. In Figure \ref{fig5}(a), the Mn and Fe magnetic moments for composition \textit{A} are shown as a function of temperature for a magnetic field of 2~T. The most striking feature is the larger magnetic moments for Mn compare to Fe. This is in line with previous neutron diffraction studies in the ferromagnetic phase.~\cite{DUN12, OU13A} This phenomenon was ascribed to the site occupancy, with Mn preferentially occupying the high moment $3g$ site and Fe the $3f$ site, which shows a weaker magnetism.~\cite{XBLIU09, HUD11, DUN12, OU13A} It is now found experimentally that this tendency is also maintained in the paramagnetic state, as was suggested by first-principle calculations.~\cite{DUN11} If we now look at the relative evolution of the projected local moments for each element, we can observe that both exhibit an abrupt decrease at \BPChem{\textit{T}\_{C}}. From 250 to 330~K, the reduction in magnetic moments is 72\% and 74\% for Fe and Mn, respectively (note that this reduction is not significantly affected by the sum-rule correction). This thus points towards a similar evolution for the magnetism of the Fe and Mn moments across the FOMT. This trend is also found in the XMCD as a function of field. Figure \ref{fig5}(b) displays the magnetic field dependence of the XMCD of composition \textit{A} in the ferromagnetic and paramagnetic states as well as at \BPChem{\textit{T}\_{C}} (in a field of 2~T). At 250~K, a field dependence characteristic for a ferromagnetic state is observed for both Fe and Mn. At 292~K, the field-induced transition can clearly be seen for both Mn and Fe. The transition is centered at a magnetic field of 2~T. In the paramagnetic state, the application of a magnetic field increases the projected magnetic moments for both Mn and Fe at an identical rate 0.06(1)~\BPChem{{$\mu$}\_{B}}$T^{-1}$. The XMCD versus \textit{T} and \textit{B} are consistent and both indicate that the Fe magnetic moment remains finite in the paramagnetic phase. The \textit{T} or \textit{B} evolution of the Fe moment is similar to that of Mn.

Let us now consider the case of composition \textit{B}, which exhibits a transitional behavior lying at the FOMT and second-order magnetic phase transition (SOMT) boundary. The XMCD as a function of temperature (at fixed field) and as a function of the magnetic field (at constant temperature) in the ferromagnetic and paramagnetic states is presented in Figure \ref{fig6}(a) and \ref{fig6}(b), respectively. The total magnetization in the ferromagnetic state is lower in composition \textit{B} than in \textit{A}. This result is in line with the previous reports in this system.~\cite{GUI14B} The Mn moments are expected to have a significantly lower magnetic moment on the $3f$ site than on the $3g$ site (and also lower than Fe on the same $3f$ site).~\cite{DUN11} This theoretical prediction is here confirmed, since the decrease in the total magnetization by the change in Mn/Fe ratio can be mainly ascribed to the reduction in Mn moment (the Mn moment is reduced more strongly than the Fe moment). Previous neutron experiments~\cite{DUN12} provided the magnetic moment on Mn in the $3g$ position (fully occupied by Mn) and their evolution with a change in Mn/Fe ratio, but failed to separate the Mn/Fe magnetic moments on the $3f$ site. The present XMCD approach allows one to obtain the Fe moment independently of the Mn signal and to estimate the Mn moment on the $3f$ site. By considering that the Mn magnetic moments on the $3g$ site are found to be independent of the Mn/Fe ratio,~\cite{DUN12, OU13A} one can derive for composition \textit{B} an experimental value for the Mn moments on the $3f$ site of \BPChem{{$\mu$}\_{Mn($3f$)}}= $\frac{1}{x}$$\langle$\BPChem{{$\mu$}\_{Mn}}$\rangle$-$\frac{1-x}{x}$\BPChem{{$\mu$}\_{Mn($3g$)}}= \BPChem{0.658~{$\mu$}\_{B}}, where $x$ is the Mn($3f$)/Mn ratio. It should be noted that (i) this moment is derived from XMCD data only and (ii) as observed for Fe moment, the Mn moment on the $3f$ site is lower than the $3g$ site. 

If one looks at the temperature evolution of the Mn/Fe moments for composition \textit{B}, once again there is a striking similarity between the respective evolution of the two elements with a reduction from 250 to 330~K of 53\% and 52\% for Fe and Mn, respectively. This Fe/Mn comparative approach for composition \textit{B} is not straightforward as the signal for Mn mixes the temperature dependence of the $3f$ and $3g$ sites. The Mn moment on the $3f$ position shows the same local moment quenching as predicted for Fe. The similarity of the Fe and Mn moments evolution however supports the overall similarity between the temperature dependence of the $3f$ and $3g$ sites observed for composition \textit{A}, \emph{i.e.}, no complete quenching of the local moment on the $3f$ site is observed for temperatures directly above the FOMT.

The similar temperature and field dependence for the Mn and Fe moments on crossing the ferro-paramagnetic transition seems to contradict the predictions from the electronic structure calculations on \BPChem{MnFeP\_{0.5}Si\_{0.5}}, for which a loss of local magnetic moment for Fe($3f$) is predicted when crossing the first-order magnetic phase transition.~\cite{DUN11} Experimentally, such an acceleration in the reduction of the $3f$ projected moments when crossing the \BPChem{\textit{T}\_{C}} is not observed. This deviation between present experimental findings and the calculations may have different causes. A first explanation for this discrepancy deals with the studied temperature range for the ferromagnetic and paramagnetic states. In the XMCD experiments, the temperatures considered are \BPChem{\textit{T}\_{C}}${\pm}$40~K. This temperature range is significantly larger than the purely discontinuous regime at the FOMT and its corresponding magnetic discontinuity. However, one is still in the transition range where short-range order develops.~\cite{FUJ88, BEC91, ZAC90, DUN12} In the paramagnetic phase, the \BPChem{\textit{M}\_{T}(\textit{B})} curves of the XMCD magnetic moments show a curvature at low magnetic field and non-perfect linearity, indicating the existence of short-range order with temporal fluctuations of ferromagnetically ordered clusters. The existence of short-range magnetic order in the paramagnetic phase has been reported in various \BPChem{Fe\_{2}P} materials up to 3\BPChem{\textit{T}\_{C}}.~\cite{FUJ88, DUN12} These fluctuations above the transition may contribute to a smearing of the Fe($3f$) quenching. Another source for the difference in Fe moment above \BPChem{\textit{T}\_{C}} may arise from the calculations themselves, like the spin arrangement used to model the paramagnetic disorder or the size of the supercell. These parameters may influence the calculated magnitude of the moment quenching for Fe($3f$) although the reduction itself seems robust for \BPChem{MnFe(Si,P)} and \BPChem{Fe\_{2}P}. To clarify the origin of this discrepancy, more work is needed both on the modeling of these materials and on the experimental side. 

\section{\label{concl}Conclusions}

The electronic and magnetic properties of \BPChem{(Mn,Fe)\_{2}(P,Si,B)} materials across their first-order magnetic phase transition have been investigated in an element-specific way by XMCD measurements. The results are compared with CT multiplet and DFT calculations. From XAS, it is found that no significant valence change and generally speaking no spectral shape modification is observed across the FOMT. From XMCD, the magnetic field and temperature dependence of the magnetic moments was obtained for the Fe and Mn moments for two Fe/Mn ratios. It is observed that the Mn exhibits a much lower magnetization on the $3f$ site than on the $3g$. In contrast to theoretical predictions, it is observed that the Fe moments are not fully quenched in the paramagnetic state. These results suggest that the magnitude of the disappearance of the $3f$ moments at \BPChem{\textit{T}\_{C}} is overestimated by ab-initio calculations. On one hand, part of this discrepancy might arise from the measurements that are done at the vicinity of the FOMT and might be smeared by the short-range order above the transition. On the theoretical side, the change in the balance between bond formation and magnetism for Fe and its neighbouring atoms might be overestimated.

\begin{acknowledgments}
We acknowledge the European Synchrotron Radiation Facility for provision of synchrotron radiation facilities and thank the beamline staff for assistance in using beam line ID08. The authors would like to thank Cinthia Piamonteze and Anne-Christine Uldry for helpful discussions. This work is financially supported by the Dutch Foundation for Fundamental Research on Matter (FOM) and BASF New Business.
\end{acknowledgments}

%\bibliography{references}% Produces the bibliography via BibTeX.

\begin{thebibliography}{57}%
\makeatletter
\providecommand \@ifxundefined [1]{%
 \@ifx{#1\undefined}
}%
\providecommand \@ifnum [1]{%
 \ifnum #1\expandafter \@firstoftwo
 \else \expandafter \@secondoftwo
 \fi
}%
\providecommand \@ifx [1]{%
 \ifx #1\expandafter \@firstoftwo
 \else \expandafter \@secondoftwo
 \fi
}%
\providecommand \natexlab [1]{#1}%
\providecommand \enquote  [1]{``#1''}%
\providecommand \bibnamefont  [1]{#1}%
\providecommand \bibfnamefont [1]{#1}%
\providecommand \citenamefont [1]{#1}%
\providecommand \href@noop [0]{\@secondoftwo}%
\providecommand \href [0]{\begingroup \@sanitize@url \@href}%
\providecommand \@href[1]{\@@startlink{#1}\@@href}%
\providecommand \@@href[1]{\endgroup#1\@@endlink}%
\providecommand \@sanitize@url [0]{\catcode `\\12\catcode `\$12\catcode
  `\&12\catcode `\#12\catcode `\^12\catcode `\_12\catcode `\%12\relax}%
\providecommand \@@startlink[1]{}%
\providecommand \@@endlink[0]{}%
\providecommand \url  [0]{\begingroup\@sanitize@url \@url }%
\providecommand \@url [1]{\endgroup\@href {#1}{\urlprefix }}%
\providecommand \urlprefix  [0]{URL }%
\providecommand \Eprint [0]{\href }%
\providecommand \doibase [0]{http://dx.doi.org/}%
\providecommand \selectlanguage [0]{\@gobble}%
\providecommand \bibinfo  [0]{\@secondoftwo}%
\providecommand \bibfield  [0]{\@secondoftwo}%
\providecommand \translation [1]{[#1]}%
\providecommand \BibitemOpen [0]{}%
\providecommand \bibitemStop [0]{}%
\providecommand \bibitemNoStop [0]{.\EOS\space}%
\providecommand \EOS [0]{\spacefactor3000\relax}%
\providecommand \BibitemShut  [1]{\csname bibitem#1\endcsname}%
\let\auto@bib@innerbib\@empty
%</preamble>
\bibitem [{\citenamefont {Tishin}\ and\ \citenamefont
  {Spichkin}(2003)}]{TIS03}%
  \BibitemOpen
  \bibfield  {author} {\bibinfo {author} {\bibfnamefont {A.~M.}\ \bibnamefont
  {Tishin}}\ and\ \bibinfo {author} {\bibfnamefont {Y.~I.}\ \bibnamefont
  {Spichkin}},\ }\href@noop {} {\emph {\bibinfo {title} {The magnetoclaoric
  effect and its Applications}}}\ (\bibinfo  {publisher} {Bristol: Institute of
  Physics Publishing},\ \bibinfo {year} {2003})\BibitemShut {NoStop}%
\bibitem [{\citenamefont {Gschneidner}\ \emph {et~al.}(2005)\citenamefont
  {Gschneidner}, \citenamefont {Pecharsky},\ and\ \citenamefont
  {Tsokol}}]{GSC05}%
  \BibitemOpen
  \bibfield  {author} {\bibinfo {author} {\bibfnamefont {K.~A.}\ \bibnamefont
  {Gschneidner}, \bibfnamefont {Jr.}}, \bibinfo {author} {\bibfnamefont
  {V.~K.}\ \bibnamefont {Pecharsky}}, \ and\ \bibinfo {author} {\bibfnamefont
  {A.~O.}\ \bibnamefont {Tsokol}},\ }\href
  {http://stacks.iop.org/0034-4885/68/i=6/a=R04} {\bibfield  {journal}
  {\bibinfo  {journal} {Rep. Prog. Phys.}\ }\textbf {\bibinfo {volume} {68}},\
  \bibinfo {pages} {1479} (\bibinfo {year} {2005})}\BibitemShut {NoStop}%
\bibitem [{\citenamefont {Smith}\ \emph {et~al.}(2012)\citenamefont {Smith},
  \citenamefont {Bahl}, \citenamefont {Bj{\o}rk}, \citenamefont {Engelbrecht},
  \citenamefont {Nielsen},\ and\ \citenamefont {Pryds}}]{SMI12}%
  \BibitemOpen
  \bibfield  {author} {\bibinfo {author} {\bibfnamefont {A.}~\bibnamefont
  {Smith}}, \bibinfo {author} {\bibfnamefont {C.~R.}\ \bibnamefont {Bahl}},
  \bibinfo {author} {\bibfnamefont {R.}~\bibnamefont {Bj{\o}rk}}, \bibinfo
  {author} {\bibfnamefont {K.}~\bibnamefont {Engelbrecht}}, \bibinfo {author}
  {\bibfnamefont {K.~K.}\ \bibnamefont {Nielsen}}, \ and\ \bibinfo {author}
  {\bibfnamefont {N.}~\bibnamefont {Pryds}},\ }\href {\doibase
  10.1002/aenm.201200167} {\bibfield  {journal} {\bibinfo  {journal} {Adv.
  Energy Mater.}\ }\textbf {\bibinfo {volume} {2}},\ \bibinfo {pages} {1288}
  (\bibinfo {year} {2012})}\BibitemShut {NoStop}%
\bibitem [{\citenamefont {Annaorazov}\ \emph {et~al.}(1996)\citenamefont
  {Annaorazov}, \citenamefont {Nikitin}, \citenamefont {Tyurin}, \citenamefont
  {Asatryan},\ and\ \citenamefont {Dovletov}}]{ANN96}%
  \BibitemOpen
  \bibfield  {author} {\bibinfo {author} {\bibfnamefont {M.~P.}\ \bibnamefont
  {Annaorazov}}, \bibinfo {author} {\bibfnamefont {S.~A.}\ \bibnamefont
  {Nikitin}}, \bibinfo {author} {\bibfnamefont {A.~L.}\ \bibnamefont {Tyurin}},
  \bibinfo {author} {\bibfnamefont {K.~A.}\ \bibnamefont {Asatryan}}, \ and\
  \bibinfo {author} {\bibfnamefont {A.~K.}\ \bibnamefont {Dovletov}},\ }\href
  {\doibase http://dx.doi.org/10.1063/1.360955} {\bibfield  {journal} {\bibinfo
   {journal} {J. Appl. Phys.}\ }\textbf {\bibinfo {volume} {79}},\ \bibinfo
  {pages} {1689} (\bibinfo {year} {1996})}\BibitemShut {NoStop}%
\bibitem [{\citenamefont {Pecharsky}\ and\ \citenamefont
  {Gschneidner}(1997)}]{PEC97}%
  \BibitemOpen
  \bibfield  {author} {\bibinfo {author} {\bibfnamefont {V.~K.}\ \bibnamefont
  {Pecharsky}}\ and\ \bibinfo {author} {\bibfnamefont {K.~A.}\ \bibnamefont
  {Gschneidner}, \bibfnamefont {Jr.}},\ }\href {\doibase
  10.1103/PhysRevLett.78.4494} {\bibfield  {journal} {\bibinfo  {journal}
  {Phys. Rev. Lett.}\ }\textbf {\bibinfo {volume} {78}},\ \bibinfo {pages}
  {4494} (\bibinfo {year} {1997})}\BibitemShut {NoStop}%
\bibitem [{\citenamefont {Wada}\ and\ \citenamefont {Tanabe}(2001)}]{WAD01}%
  \BibitemOpen
  \bibfield  {author} {\bibinfo {author} {\bibfnamefont {H.}~\bibnamefont
  {Wada}}\ and\ \bibinfo {author} {\bibfnamefont {Y.}~\bibnamefont {Tanabe}},\
  }\href {\doibase http://dx.doi.org/10.1063/1.1419048} {\bibfield  {journal}
  {\bibinfo  {journal} {Appl. Phys. Lett.}\ }\textbf {\bibinfo {volume} {79}},\
  \bibinfo {pages} {3302} (\bibinfo {year} {2001})}\BibitemShut {NoStop}%
\bibitem [{\citenamefont {Hu}\ \emph {et~al.}(2001)\citenamefont {Hu},
  \citenamefont {Shen}, \citenamefont {Sun}, \citenamefont {Cheng},
  \citenamefont {Rao},\ and\ \citenamefont {Zhang}}]{Hu01}%
  \BibitemOpen
  \bibfield  {author} {\bibinfo {author} {\bibfnamefont {F.-X.}\ \bibnamefont
  {Hu}}, \bibinfo {author} {\bibfnamefont {B.-G.}\ \bibnamefont {Shen}},
  \bibinfo {author} {\bibfnamefont {J.-R.}\ \bibnamefont {Sun}}, \bibinfo
  {author} {\bibfnamefont {Z.-H.}\ \bibnamefont {Cheng}}, \bibinfo {author}
  {\bibfnamefont {G.-H.}\ \bibnamefont {Rao}}, \ and\ \bibinfo {author}
  {\bibfnamefont {X.-X.}\ \bibnamefont {Zhang}},\ }\href {\doibase
  http://dx.doi.org/10.1063/1.1375836} {\bibfield  {journal} {\bibinfo
  {journal} {Appl. Phys. Lett.}\ }\textbf {\bibinfo {volume} {78}},\ \bibinfo
  {pages} {3675} (\bibinfo {year} {2001})}\BibitemShut {NoStop}%
\bibitem [{\citenamefont {Fujita}\ \emph {et~al.}(2003)\citenamefont {Fujita},
  \citenamefont {Fujieda}, \citenamefont {Hasegawa},\ and\ \citenamefont
  {Fukamichi}}]{FUJ03}%
  \BibitemOpen
  \bibfield  {author} {\bibinfo {author} {\bibfnamefont {A.}~\bibnamefont
  {Fujita}}, \bibinfo {author} {\bibfnamefont {S.}~\bibnamefont {Fujieda}},
  \bibinfo {author} {\bibfnamefont {Y.}~\bibnamefont {Hasegawa}}, \ and\
  \bibinfo {author} {\bibfnamefont {K.}~\bibnamefont {Fukamichi}},\ }\href
  {\doibase 10.1103/PhysRevB.67.104416} {\bibfield  {journal} {\bibinfo
  {journal} {Phys. Rev. B}\ }\textbf {\bibinfo {volume} {67}},\ \bibinfo
  {pages} {104416} (\bibinfo {year} {2003})}\BibitemShut {NoStop}%
\bibitem [{\citenamefont {Tegus}\ \emph {et~al.}(2002)\citenamefont {Tegus},
  \citenamefont {Br\"uck}, \citenamefont {Buschow},\ and\ \citenamefont
  {de~Boer}}]{TEG02}%
  \BibitemOpen
  \bibfield  {author} {\bibinfo {author} {\bibfnamefont {O.}~\bibnamefont
  {Tegus}}, \bibinfo {author} {\bibfnamefont {E.}~\bibnamefont {Br\"uck}},
  \bibinfo {author} {\bibfnamefont {K.~H.~J.}\ \bibnamefont {Buschow}}, \ and\
  \bibinfo {author} {\bibfnamefont {F.~R.}\ \bibnamefont {de~Boer}},\ }\href
  {\doibase 10.1038/415150a} {\bibfield  {journal} {\bibinfo  {journal}
  {Nature}\ }\textbf {\bibinfo {volume} {415}},\ \bibinfo {pages} {150}
  (\bibinfo {year} {2002})}\BibitemShut {NoStop}%
\bibitem [{\citenamefont {Trung}\ \emph {et~al.}(2010)\citenamefont {Trung},
  \citenamefont {Zhang}, \citenamefont {Caron}, \citenamefont {Buschow},\ and\
  \citenamefont {Br\"uck}}]{TRU10}%
  \BibitemOpen
  \bibfield  {author} {\bibinfo {author} {\bibfnamefont {N.~T.}\ \bibnamefont
  {Trung}}, \bibinfo {author} {\bibfnamefont {L.}~\bibnamefont {Zhang}},
  \bibinfo {author} {\bibfnamefont {L.}~\bibnamefont {Caron}}, \bibinfo
  {author} {\bibfnamefont {K.~H.~J.}\ \bibnamefont {Buschow}}, \ and\ \bibinfo
  {author} {\bibfnamefont {E.}~\bibnamefont {Br\"uck}},\ }\href {\doibase
  http://dx.doi.org/10.1063/1.3399773} {\bibfield  {journal} {\bibinfo
  {journal} {Appl. Phys. Lett.}\ }\textbf {\bibinfo {volume} {96}},\ \bibinfo
  {pages} {172540} (\bibinfo {year} {2010})}\BibitemShut {NoStop}%
\bibitem [{\citenamefont {Dung}\ \emph {et~al.}(2011)\citenamefont {Dung},
  \citenamefont {Ou}, \citenamefont {Caron}, \citenamefont {Zhang},
  \citenamefont {Thanh}, \citenamefont {de~Wijs}, \citenamefont {de~Groot},
  \citenamefont {Buschow},\ and\ \citenamefont {Br\"uck}}]{DUN11}%
  \BibitemOpen
  \bibfield  {author} {\bibinfo {author} {\bibfnamefont {N.~H.}\ \bibnamefont
  {Dung}}, \bibinfo {author} {\bibfnamefont {Z.~Q.}\ \bibnamefont {Ou}},
  \bibinfo {author} {\bibfnamefont {L.}~\bibnamefont {Caron}}, \bibinfo
  {author} {\bibfnamefont {L.}~\bibnamefont {Zhang}}, \bibinfo {author}
  {\bibfnamefont {D.~T.~C.}\ \bibnamefont {Thanh}}, \bibinfo {author}
  {\bibfnamefont {G.~A.}\ \bibnamefont {de~Wijs}}, \bibinfo {author}
  {\bibfnamefont {R.~A.}\ \bibnamefont {de~Groot}}, \bibinfo {author}
  {\bibfnamefont {K.~H.~J.}\ \bibnamefont {Buschow}}, \ and\ \bibinfo {author}
  {\bibfnamefont {E.}~\bibnamefont {Br\"uck}},\ }\href {\doibase
  10.1002/aenm.201100252} {\bibfield  {journal} {\bibinfo  {journal} {Adv.
  Energy Mater.}\ }\textbf {\bibinfo {volume} {1}},\ \bibinfo {pages} {1215}
  (\bibinfo {year} {2011})}\BibitemShut {NoStop}%
\bibitem [{\citenamefont {Yibole}\ \emph {et~al.}(2014)\citenamefont {Yibole},
  \citenamefont {Guillou}, \citenamefont {Zhang}, \citenamefont {van Dijk},\
  and\ \citenamefont {Br\"uck}}]{YIB14}%
  \BibitemOpen
  \bibfield  {author} {\bibinfo {author} {\bibfnamefont {H.}~\bibnamefont
  {Yibole}}, \bibinfo {author} {\bibfnamefont {F.}~\bibnamefont {Guillou}},
  \bibinfo {author} {\bibfnamefont {L.}~\bibnamefont {Zhang}}, \bibinfo
  {author} {\bibfnamefont {N.~H.}\ \bibnamefont {van Dijk}}, \ and\ \bibinfo
  {author} {\bibfnamefont {E.}~\bibnamefont {Br\"uck}},\ }\href
  {http://stacks.iop.org/0022-3727/47/i=7/a=075002} {\bibfield  {journal}
  {\bibinfo  {journal} {J. Phys. D: Appl. Phys.}\ }\textbf {\bibinfo {volume}
  {47}},\ \bibinfo {pages} {075002} (\bibinfo {year} {2014})}\BibitemShut
  {NoStop}%
\bibitem [{\citenamefont {Guillou}\ \emph
  {et~al.}(2014{\natexlab{a}})\citenamefont {Guillou}, \citenamefont {Porcari},
  \citenamefont {Yibole}, \citenamefont {van Dijk},\ and\ \citenamefont
  {Br\"uck}}]{GUI14}%
  \BibitemOpen
  \bibfield  {author} {\bibinfo {author} {\bibfnamefont {F.}~\bibnamefont
  {Guillou}}, \bibinfo {author} {\bibfnamefont {G.}~\bibnamefont {Porcari}},
  \bibinfo {author} {\bibfnamefont {H.}~\bibnamefont {Yibole}}, \bibinfo
  {author} {\bibfnamefont {N.}~\bibnamefont {van Dijk}}, \ and\ \bibinfo
  {author} {\bibfnamefont {E.}~\bibnamefont {Br\"uck}},\ }\href {\doibase
  10.1002/adma.201304788} {\bibfield  {journal} {\bibinfo  {journal} {Adv.
  Mater.}\ }\textbf {\bibinfo {volume} {26}},\ \bibinfo {pages} {2671}
  (\bibinfo {year} {2014}{\natexlab{a}})}\BibitemShut {NoStop}%
\bibitem [{\citenamefont {Pecharsky}\ \emph {et~al.}(2001)\citenamefont
  {Pecharsky}, \citenamefont {Gschneidner}, \citenamefont {Pecharsky},\ and\
  \citenamefont {Tishin}}]{PER01}%
  \BibitemOpen
  \bibfield  {author} {\bibinfo {author} {\bibfnamefont {V.~K.}\ \bibnamefont
  {Pecharsky}}, \bibinfo {author} {\bibfnamefont {K.~A.}\ \bibnamefont
  {Gschneidner}}, \bibinfo {author} {\bibfnamefont {A.~O.}\ \bibnamefont
  {Pecharsky}}, \ and\ \bibinfo {author} {\bibfnamefont {A.~M.}\ \bibnamefont
  {Tishin}},\ }\href {\doibase 10.1103/PhysRevB.64.144406} {\bibfield
  {journal} {\bibinfo  {journal} {Phys. Rev. B}\ }\textbf {\bibinfo {volume}
  {64}},\ \bibinfo {pages} {144406} (\bibinfo {year} {2001})}\BibitemShut
  {NoStop}%
\bibitem [{\citenamefont {Annaorazov}\ \emph {et~al.}(2004)\citenamefont
  {Annaorazov}, \citenamefont {B\"arner},\ and\ \citenamefont
  {Yal\c{c}in}}]{ANN04}%
  \BibitemOpen
  \bibfield  {author} {\bibinfo {author} {\bibfnamefont {M.}~\bibnamefont
  {Annaorazov}}, \bibinfo {author} {\bibfnamefont {K.}~\bibnamefont
  {B\"arner}}, \ and\ \bibinfo {author} {\bibfnamefont {.}~\bibnamefont
  {Yal\c{c}in}},\ }\href {\doibase
  http://dx.doi.org/10.1016/j.jallcom.2003.10.012} {\bibfield  {journal}
  {\bibinfo  {journal} {J. Alloys Compd.}\ }\textbf {\bibinfo {volume} {372}},\
  \bibinfo {pages} {52 } (\bibinfo {year} {2004})}\BibitemShut {NoStop}%
\bibitem [{\citenamefont {Sandeman}(2012)}]{SAN12}%
  \BibitemOpen
  \bibfield  {author} {\bibinfo {author} {\bibfnamefont {K.~G.}\ \bibnamefont
  {Sandeman}},\ }\href {\doibase
  http://dx.doi.org/10.1016/j.scriptamat.2012.02.045} {\bibfield  {journal}
  {\bibinfo  {journal} {Scripta Mater.}\ }\textbf {\bibinfo {volume} {67}},\
  \bibinfo {pages} {566 } (\bibinfo {year} {2012})}\BibitemShut {NoStop}%
\bibitem [{\citenamefont {Beckman}\ and\ \citenamefont
  {Lundgren}(1991)}]{BEC91}%
  \BibitemOpen
  \bibfield  {author} {\bibinfo {author} {\bibfnamefont {O.}~\bibnamefont
  {Beckman}}\ and\ \bibinfo {author} {\bibfnamefont {L.}~\bibnamefont
  {Lundgren}},\ }\href@noop {} {\emph {\bibinfo {title} {Handbook of Magnetic
  materials}}}\ (\bibinfo  {publisher} {North-Holland, Amsterdam},\ \bibinfo
  {year} {1991})\BibitemShut {NoStop}%
\bibitem [{\citenamefont {Dung}\ \emph
  {et~al.}(2012{\natexlab{a}})\citenamefont {Dung}, \citenamefont {Zhang},
  \citenamefont {Ou}, \citenamefont {Zhao}, \citenamefont {van Eijck},
  \citenamefont {Mulders}, \citenamefont {Avdeev}, \citenamefont {Suard},
  \citenamefont {van Dijk},\ and\ \citenamefont {Br\"uck}}]{DUN12}%
  \BibitemOpen
  \bibfield  {author} {\bibinfo {author} {\bibfnamefont {N.~H.}\ \bibnamefont
  {Dung}}, \bibinfo {author} {\bibfnamefont {L.}~\bibnamefont {Zhang}},
  \bibinfo {author} {\bibfnamefont {Z.~Q.}\ \bibnamefont {Ou}}, \bibinfo
  {author} {\bibfnamefont {L.}~\bibnamefont {Zhao}}, \bibinfo {author}
  {\bibfnamefont {L.}~\bibnamefont {van Eijck}}, \bibinfo {author}
  {\bibfnamefont {A.~M.}\ \bibnamefont {Mulders}}, \bibinfo {author}
  {\bibfnamefont {M.}~\bibnamefont {Avdeev}}, \bibinfo {author} {\bibfnamefont
  {E.}~\bibnamefont {Suard}}, \bibinfo {author} {\bibfnamefont {N.~H.}\
  \bibnamefont {van Dijk}}, \ and\ \bibinfo {author} {\bibfnamefont
  {E.}~\bibnamefont {Br\"uck}},\ }\href {\doibase 10.1103/PhysRevB.86.045134}
  {\bibfield  {journal} {\bibinfo  {journal} {Phys. Rev. B}\ }\textbf {\bibinfo
  {volume} {86}},\ \bibinfo {pages} {045134} (\bibinfo {year}
  {2012}{\natexlab{a}})}\BibitemShut {NoStop}%
\bibitem [{\citenamefont {Zhang}\ \emph {et~al.}(2005)\citenamefont {Zhang},
  \citenamefont {Mo\v{z}e}, \citenamefont {Proke\v{s}}, \citenamefont {Tegus},\
  and\ \citenamefont {Br\"uck}}]{LIA05}%
  \BibitemOpen
  \bibfield  {author} {\bibinfo {author} {\bibfnamefont {L.}~\bibnamefont
  {Zhang}}, \bibinfo {author} {\bibfnamefont {O.}~\bibnamefont {Mo\v{z}e}},
  \bibinfo {author} {\bibfnamefont {K.}~\bibnamefont {Proke\v{s}}}, \bibinfo
  {author} {\bibfnamefont {O.}~\bibnamefont {Tegus}}, \ and\ \bibinfo {author}
  {\bibfnamefont {E.}~\bibnamefont {Br\"uck}},\ }\href {\doibase
  http://dx.doi.org/10.1016/j.jmmm.2004.11.335} {\bibfield  {journal} {\bibinfo
   {journal} {J. Magn. Magn. Mater.}\ }\textbf {\bibinfo {volume} {290}},\
  \bibinfo {pages} {679 } (\bibinfo {year} {2005})}\BibitemShut {NoStop}%
\bibitem [{\citenamefont {Liu}\ \emph {et~al.}(2009)\citenamefont {Liu},
  \citenamefont {Huang}, \citenamefont {Yue}, \citenamefont {Lynn},
  \citenamefont {Liu}, \citenamefont {Chen}, \citenamefont {Wu},\ and\
  \citenamefont {Zhang}}]{LIU09}%
  \BibitemOpen
  \bibfield  {author} {\bibinfo {author} {\bibfnamefont {D.~M.}\ \bibnamefont
  {Liu}}, \bibinfo {author} {\bibfnamefont {Q.~Z.}\ \bibnamefont {Huang}},
  \bibinfo {author} {\bibfnamefont {M.}~\bibnamefont {Yue}}, \bibinfo {author}
  {\bibfnamefont {J.~W.}\ \bibnamefont {Lynn}}, \bibinfo {author}
  {\bibfnamefont {L.~J.}\ \bibnamefont {Liu}}, \bibinfo {author} {\bibfnamefont
  {Y.}~\bibnamefont {Chen}}, \bibinfo {author} {\bibfnamefont {Z.~H.}\
  \bibnamefont {Wu}}, \ and\ \bibinfo {author} {\bibfnamefont {J.~X.}\
  \bibnamefont {Zhang}},\ }\href {\doibase 10.1103/PhysRevB.80.174415}
  {\bibfield  {journal} {\bibinfo  {journal} {Phys. Rev. B}\ }\textbf {\bibinfo
  {volume} {80}},\ \bibinfo {pages} {174415} (\bibinfo {year}
  {2009})}\BibitemShut {NoStop}%
\bibitem [{\citenamefont {H\"{o}glin}\ \emph {et~al.}(2011)\citenamefont
  {H\"{o}glin}, \citenamefont {Hudl}, \citenamefont {Sahlberg}, \citenamefont
  {Nordblad}, \citenamefont {Beran},\ and\ \citenamefont {Andersson}}]{HOG11}%
  \BibitemOpen
  \bibfield  {author} {\bibinfo {author} {\bibfnamefont {V.}~\bibnamefont
  {H\"{o}glin}}, \bibinfo {author} {\bibfnamefont {M.}~\bibnamefont {Hudl}},
  \bibinfo {author} {\bibfnamefont {M.}~\bibnamefont {Sahlberg}}, \bibinfo
  {author} {\bibfnamefont {P.}~\bibnamefont {Nordblad}}, \bibinfo {author}
  {\bibfnamefont {P.}~\bibnamefont {Beran}}, \ and\ \bibinfo {author}
  {\bibfnamefont {Y.}~\bibnamefont {Andersson}},\ }\href {\doibase
  http://dx.doi.org/10.1016/j.jssc.2011.06.019} {\bibfield  {journal} {\bibinfo
   {journal} {J. Solid State Chem.}\ }\textbf {\bibinfo {volume} {184}},\
  \bibinfo {pages} {2434 } (\bibinfo {year} {2011})}\BibitemShut {NoStop}%
\bibitem [{\citenamefont {Ou}\ \emph {et~al.}(2013)\citenamefont {Ou},
  \citenamefont {Zhang}, \citenamefont {Dung}, \citenamefont {van Eijck},
  \citenamefont {Mulders}, \citenamefont {Avdeev}, \citenamefont {van Dijk},\
  and\ \citenamefont {Br\"uck}}]{OU13A}%
  \BibitemOpen
  \bibfield  {author} {\bibinfo {author} {\bibfnamefont {Z.}~\bibnamefont
  {Ou}}, \bibinfo {author} {\bibfnamefont {L.}~\bibnamefont {Zhang}}, \bibinfo
  {author} {\bibfnamefont {N.}~\bibnamefont {Dung}}, \bibinfo {author}
  {\bibfnamefont {L.}~\bibnamefont {van Eijck}}, \bibinfo {author}
  {\bibfnamefont {A.}~\bibnamefont {Mulders}}, \bibinfo {author} {\bibfnamefont
  {M.}~\bibnamefont {Avdeev}}, \bibinfo {author} {\bibfnamefont
  {N.}~\bibnamefont {van Dijk}}, \ and\ \bibinfo {author} {\bibfnamefont
  {E.}~\bibnamefont {Br\"uck}},\ }\href {\doibase
  http://dx.doi.org/10.1016/j.jmmm.2013.03.028} {\bibfield  {journal} {\bibinfo
   {journal} {J. Magn. Magn. Mater.}\ }\textbf {\bibinfo {volume} {340}},\
  \bibinfo {pages} {80 } (\bibinfo {year} {2013})}\BibitemShut {NoStop}%
\bibitem [{\citenamefont {Yue}\ \emph {et~al.}(2013)\citenamefont {Yue},
  \citenamefont {Liu}, \citenamefont {Huang}, \citenamefont {Wang},
  \citenamefont {Hu}, \citenamefont {Li}, \citenamefont {Rao}, \citenamefont
  {Shen}, \citenamefont {Lynn},\ and\ \citenamefont {Zhang}}]{YUE13}%
  \BibitemOpen
  \bibfield  {author} {\bibinfo {author} {\bibfnamefont {M.}~\bibnamefont
  {Yue}}, \bibinfo {author} {\bibfnamefont {D.}~\bibnamefont {Liu}}, \bibinfo
  {author} {\bibfnamefont {Q.}~\bibnamefont {Huang}}, \bibinfo {author}
  {\bibfnamefont {T.}~\bibnamefont {Wang}}, \bibinfo {author} {\bibfnamefont
  {F.}~\bibnamefont {Hu}}, \bibinfo {author} {\bibfnamefont {J.}~\bibnamefont
  {Li}}, \bibinfo {author} {\bibfnamefont {G.}~\bibnamefont {Rao}}, \bibinfo
  {author} {\bibfnamefont {B.}~\bibnamefont {Shen}}, \bibinfo {author}
  {\bibfnamefont {J.~W.}\ \bibnamefont {Lynn}}, \ and\ \bibinfo {author}
  {\bibfnamefont {J.}~\bibnamefont {Zhang}},\ }\href {\doibase
  http://dx.doi.org/10.1063/1.4788803} {\bibfield  {journal} {\bibinfo
  {journal} {J. Appl. Phys.}\ }\textbf {\bibinfo {volume} {113}},\ \bibinfo
  {eid} {043925} (\bibinfo {year} {2013})}\BibitemShut {NoStop}%
\bibitem [{\citenamefont {Yamada}\ and\ \citenamefont {Terao}(2002)}]{YAM02}%
  \BibitemOpen
  \bibfield  {author} {\bibinfo {author} {\bibfnamefont {H.}~\bibnamefont
  {Yamada}}\ and\ \bibinfo {author} {\bibfnamefont {K.}~\bibnamefont {Terao}},\
  }\href {\doibase 10.1080/01411590290023120} {\bibfield  {journal} {\bibinfo
  {journal} {Phase Transitions}\ }\textbf {\bibinfo {volume} {75}},\ \bibinfo
  {pages} {231} (\bibinfo {year} {2002})}\BibitemShut {NoStop}%
\bibitem [{\citenamefont {Gercsi}\ \emph {et~al.}(2013)\citenamefont {Gercsi},
  \citenamefont {Delczeg-Czirjak}, \citenamefont {Vitos}, \citenamefont
  {Wills}, \citenamefont {Daoud-Aladine},\ and\ \citenamefont
  {Sandeman}}]{GER13}%
  \BibitemOpen
  \bibfield  {author} {\bibinfo {author} {\bibfnamefont {Z.}~\bibnamefont
  {Gercsi}}, \bibinfo {author} {\bibfnamefont {E.~K.}\ \bibnamefont
  {Delczeg-Czirjak}}, \bibinfo {author} {\bibfnamefont {L.}~\bibnamefont
  {Vitos}}, \bibinfo {author} {\bibfnamefont {A.~S.}\ \bibnamefont {Wills}},
  \bibinfo {author} {\bibfnamefont {A.}~\bibnamefont {Daoud-Aladine}}, \ and\
  \bibinfo {author} {\bibfnamefont {K.~G.}\ \bibnamefont {Sandeman}},\ }\href
  {\doibase 10.1103/PhysRevB.88.024417} {\bibfield  {journal} {\bibinfo
  {journal} {Phys. Rev. B}\ }\textbf {\bibinfo {volume} {88}},\ \bibinfo
  {pages} {024417} (\bibinfo {year} {2013})}\BibitemShut {NoStop}%
\bibitem [{\citenamefont {Dung}\ \emph
  {et~al.}(2012{\natexlab{b}})\citenamefont {Dung}, \citenamefont {Zhang},
  \citenamefont {Ou},\ and\ \citenamefont {Br\"uck}}]{DUN12B}%
  \BibitemOpen
  \bibfield  {author} {\bibinfo {author} {\bibfnamefont {N.~H.}\ \bibnamefont
  {Dung}}, \bibinfo {author} {\bibfnamefont {L.}~\bibnamefont {Zhang}},
  \bibinfo {author} {\bibfnamefont {Z.~Q.}\ \bibnamefont {Ou}}, \ and\ \bibinfo
  {author} {\bibfnamefont {E.}~\bibnamefont {Br\"uck}},\ }\href {\doibase
  http://dx.doi.org/10.1016/j.scriptamat.2012.08.036} {\bibfield  {journal}
  {\bibinfo  {journal} {Scripta Mater.}\ }\textbf {\bibinfo {volume} {67}},\
  \bibinfo {pages} {975 } (\bibinfo {year} {2012}{\natexlab{b}})}\BibitemShut
  {NoStop}%
\bibitem [{\citenamefont {de~Groot}\ \emph {et~al.}(1990)\citenamefont
  {de~Groot}, \citenamefont {Fuggle}, \citenamefont {Thole},\ and\
  \citenamefont {Sawatzky}}]{GRO90}%
  \BibitemOpen
  \bibfield  {author} {\bibinfo {author} {\bibfnamefont {F.~M.~F.}\
  \bibnamefont {de~Groot}}, \bibinfo {author} {\bibfnamefont {J.~C.}\
  \bibnamefont {Fuggle}}, \bibinfo {author} {\bibfnamefont {B.~T.}\
  \bibnamefont {Thole}}, \ and\ \bibinfo {author} {\bibfnamefont {G.~A.}\
  \bibnamefont {Sawatzky}},\ }\href {\doibase 10.1103/PhysRevB.42.5459}
  {\bibfield  {journal} {\bibinfo  {journal} {Phys. Rev. B}\ }\textbf {\bibinfo
  {volume} {42}},\ \bibinfo {pages} {5459} (\bibinfo {year}
  {1990})}\BibitemShut {NoStop}%
\bibitem [{\citenamefont {de~Groot}(2005)}]{GRO05}%
  \BibitemOpen
  \bibfield  {author} {\bibinfo {author} {\bibfnamefont {F.}~\bibnamefont
  {de~Groot}},\ }\href {\doibase http://dx.doi.org/10.1016/j.ccr.2004.03.018}
  {\bibfield  {journal} {\bibinfo  {journal} {Coord. Chem. Rev.}\ }\textbf
  {\bibinfo {volume} {249}},\ \bibinfo {pages} {31 } (\bibinfo {year}
  {2005})}\BibitemShut {NoStop}%
\bibitem [{\citenamefont {Stavitski}\ and\ \citenamefont
  {de~Groot}(2010)}]{STA10}%
  \BibitemOpen
  \bibfield  {author} {\bibinfo {author} {\bibfnamefont {E.}~\bibnamefont
  {Stavitski}}\ and\ \bibinfo {author} {\bibfnamefont {F.~M.}\ \bibnamefont
  {de~Groot}},\ }\href {\doibase
  http://dx.doi.org/10.1016/j.micron.2010.06.005} {\bibfield  {journal}
  {\bibinfo  {journal} {Micron}\ }\textbf {\bibinfo {volume} {41}},\ \bibinfo
  {pages} {687 } (\bibinfo {year} {2010})}\BibitemShut {NoStop}%
\bibitem [{\citenamefont {Blaha}\ \emph {et~al.}(2001)\citenamefont {Blaha},
  \citenamefont {Schwarz}, \citenamefont {Madsen}, \citenamefont {Kvasnicka},\
  and\ \citenamefont {Luitz}}]{BLA01}%
  \BibitemOpen
  \bibfield  {author} {\bibinfo {author} {\bibfnamefont {P.}~\bibnamefont
  {Blaha}}, \bibinfo {author} {\bibfnamefont {K.}~\bibnamefont {Schwarz}},
  \bibinfo {author} {\bibfnamefont {G.}~\bibnamefont {Madsen}}, \bibinfo
  {author} {\bibfnamefont {D.}~\bibnamefont {Kvasnicka}}, \ and\ \bibinfo
  {author} {\bibfnamefont {J.}~\bibnamefont {Luitz}},\ }\href@noop {} {\emph
  {\bibinfo {title} {WIEN2k}}}\ (\bibinfo  {publisher} {TU Wien, Austria},\
  \bibinfo {year} {2001})\BibitemShut {NoStop}%
\bibitem [{\citenamefont {Singh}(1994)}]{SIN94}%
  \BibitemOpen
  \bibfield  {author} {\bibinfo {author} {\bibfnamefont {D.}~\bibnamefont
  {Singh}},\ }\href@noop {} {\emph {\bibinfo {title} {Planewaves,
  pseudopotentials and the lapw method}}}\ (\bibinfo  {publisher} {Kluwer
  Academic Publishers, London},\ \bibinfo {year} {1994})\BibitemShut {NoStop}%
\bibitem [{\citenamefont {Perdew}\ \emph {et~al.}(1996)\citenamefont {Perdew},
  \citenamefont {Burke},\ and\ \citenamefont {Ernzerhof}}]{PER96}%
  \BibitemOpen
  \bibfield  {author} {\bibinfo {author} {\bibfnamefont {J.~P.}\ \bibnamefont
  {Perdew}}, \bibinfo {author} {\bibfnamefont {K.}~\bibnamefont {Burke}}, \
  and\ \bibinfo {author} {\bibfnamefont {M.}~\bibnamefont {Ernzerhof}},\ }\href
  {\doibase 10.1103/PhysRevLett.77.3865} {\bibfield  {journal} {\bibinfo
  {journal} {Phys. Rev. Lett.}\ }\textbf {\bibinfo {volume} {77}},\ \bibinfo
  {pages} {3865} (\bibinfo {year} {1996})}\BibitemShut {NoStop}%
\bibitem [{\citenamefont {Ambrosch-Draxl}\ and\ \citenamefont
  {Sofo}(2006)}]{AMB06}%
  \BibitemOpen
  \bibfield  {author} {\bibinfo {author} {\bibfnamefont {C.}~\bibnamefont
  {Ambrosch-Draxl}}\ and\ \bibinfo {author} {\bibfnamefont {J.~O.}\
  \bibnamefont {Sofo}},\ }\href {\doibase
  http://dx.doi.org/10.1016/j.cpc.2006.03.005} {\bibfield  {journal} {\bibinfo
  {journal} {Comp. Phys. Comm.}\ }\textbf {\bibinfo {volume} {175}},\ \bibinfo
  {pages} {1 } (\bibinfo {year} {2006})}\BibitemShut {NoStop}%
\bibitem [{\citenamefont {Edmonds}\ \emph {et~al.}(2006)\citenamefont
  {Edmonds}, \citenamefont {van~der Laan}, \citenamefont {Freeman},
  \citenamefont {Farley}, \citenamefont {Johal}, \citenamefont {Campion},
  \citenamefont {Foxon}, \citenamefont {Gallagher},\ and\ \citenamefont
  {Arenholz}}]{EDM06}%
  \BibitemOpen
  \bibfield  {author} {\bibinfo {author} {\bibfnamefont {K.~W.}\ \bibnamefont
  {Edmonds}}, \bibinfo {author} {\bibfnamefont {G.}~\bibnamefont {van~der
  Laan}}, \bibinfo {author} {\bibfnamefont {A.~A.}\ \bibnamefont {Freeman}},
  \bibinfo {author} {\bibfnamefont {N.~R.~S.}\ \bibnamefont {Farley}}, \bibinfo
  {author} {\bibfnamefont {T.~K.}\ \bibnamefont {Johal}}, \bibinfo {author}
  {\bibfnamefont {R.~P.}\ \bibnamefont {Campion}}, \bibinfo {author}
  {\bibfnamefont {C.~T.}\ \bibnamefont {Foxon}}, \bibinfo {author}
  {\bibfnamefont {B.~L.}\ \bibnamefont {Gallagher}}, \ and\ \bibinfo {author}
  {\bibfnamefont {E.}~\bibnamefont {Arenholz}},\ }\href {\doibase
  10.1103/PhysRevLett.96.117207} {\bibfield  {journal} {\bibinfo  {journal}
  {Phys. Rev. Lett.}\ }\textbf {\bibinfo {volume} {96}},\ \bibinfo {pages}
  {117207} (\bibinfo {year} {2006})}\BibitemShut {NoStop}%
\bibitem [{\citenamefont {Klaer}\ \emph {et~al.}(2011)\citenamefont {Klaer},
  \citenamefont {Balke}, \citenamefont {Alijani}, \citenamefont {Winterlik},
  \citenamefont {Fecher}, \citenamefont {Felser},\ and\ \citenamefont
  {Elmers}}]{KLA11}%
  \BibitemOpen
  \bibfield  {author} {\bibinfo {author} {\bibfnamefont {P.}~\bibnamefont
  {Klaer}}, \bibinfo {author} {\bibfnamefont {B.}~\bibnamefont {Balke}},
  \bibinfo {author} {\bibfnamefont {V.}~\bibnamefont {Alijani}}, \bibinfo
  {author} {\bibfnamefont {J.}~\bibnamefont {Winterlik}}, \bibinfo {author}
  {\bibfnamefont {G.~H.}\ \bibnamefont {Fecher}}, \bibinfo {author}
  {\bibfnamefont {C.}~\bibnamefont {Felser}}, \ and\ \bibinfo {author}
  {\bibfnamefont {H.~J.}\ \bibnamefont {Elmers}},\ }\href {\doibase
  10.1103/PhysRevB.84.144413} {\bibfield  {journal} {\bibinfo  {journal} {Phys.
  Rev. B}\ }\textbf {\bibinfo {volume} {84}},\ \bibinfo {pages} {144413}
  (\bibinfo {year} {2011})}\BibitemShut {NoStop}%
\bibitem [{\citenamefont {Priolkar}\ \emph {et~al.}(2013)\citenamefont
  {Priolkar}, \citenamefont {Bhobe}, \citenamefont {Lobo}, \citenamefont
  {D'Souza}, \citenamefont {Barman}, \citenamefont {Chakrabarti},\ and\
  \citenamefont {Emura}}]{PRI13}%
  \BibitemOpen
  \bibfield  {author} {\bibinfo {author} {\bibfnamefont {K.~R.}\ \bibnamefont
  {Priolkar}}, \bibinfo {author} {\bibfnamefont {P.~A.}\ \bibnamefont {Bhobe}},
  \bibinfo {author} {\bibfnamefont {D.~N.}\ \bibnamefont {Lobo}}, \bibinfo
  {author} {\bibfnamefont {S.~W.}\ \bibnamefont {D'Souza}}, \bibinfo {author}
  {\bibfnamefont {S.~R.}\ \bibnamefont {Barman}}, \bibinfo {author}
  {\bibfnamefont {A.}~\bibnamefont {Chakrabarti}}, \ and\ \bibinfo {author}
  {\bibfnamefont {S.}~\bibnamefont {Emura}},\ }\href {\doibase
  10.1103/PhysRevB.87.144412} {\bibfield  {journal} {\bibinfo  {journal} {Phys.
  Rev. B}\ }\textbf {\bibinfo {volume} {87}},\ \bibinfo {pages} {144412}
  (\bibinfo {year} {2013})}\BibitemShut {NoStop}%
\bibitem [{\citenamefont {D\"urr}\ \emph {et~al.}(1997)\citenamefont {D\"urr},
  \citenamefont {van~der Laan}, \citenamefont {Spanke}, \citenamefont
  {Hillebrecht},\ and\ \citenamefont {Brookes}}]{DUR97}%
  \BibitemOpen
  \bibfield  {author} {\bibinfo {author} {\bibfnamefont {H.~A.}\ \bibnamefont
  {D\"urr}}, \bibinfo {author} {\bibfnamefont {G.}~\bibnamefont {van~der
  Laan}}, \bibinfo {author} {\bibfnamefont {D.}~\bibnamefont {Spanke}},
  \bibinfo {author} {\bibfnamefont {F.~U.}\ \bibnamefont {Hillebrecht}}, \ and\
  \bibinfo {author} {\bibfnamefont {N.~B.}\ \bibnamefont {Brookes}},\ }\href
  {\doibase 10.1103/PhysRevB.56.8156} {\bibfield  {journal} {\bibinfo
  {journal} {Phys. Rev. B}\ }\textbf {\bibinfo {volume} {56}},\ \bibinfo
  {pages} {8156} (\bibinfo {year} {1997})}\BibitemShut {NoStop}%
\bibitem [{\citenamefont {Caron}\ \emph {et~al.}(2013)\citenamefont {Caron},
  \citenamefont {Hudl}, \citenamefont {H\"oglin}, \citenamefont {Dung},
  \citenamefont {Gomez}, \citenamefont {Sahlberg}, \citenamefont {Br\"uck},
  \citenamefont {Andersson},\ and\ \citenamefont {Nordblad}}]{CAR14}%
  \BibitemOpen
  \bibfield  {author} {\bibinfo {author} {\bibfnamefont {L.}~\bibnamefont
  {Caron}}, \bibinfo {author} {\bibfnamefont {M.}~\bibnamefont {Hudl}},
  \bibinfo {author} {\bibfnamefont {V.}~\bibnamefont {H\"oglin}}, \bibinfo
  {author} {\bibfnamefont {N.~H.}\ \bibnamefont {Dung}}, \bibinfo {author}
  {\bibfnamefont {C.~P.}\ \bibnamefont {Gomez}}, \bibinfo {author}
  {\bibfnamefont {M.}~\bibnamefont {Sahlberg}}, \bibinfo {author}
  {\bibfnamefont {E.}~\bibnamefont {Br\"uck}}, \bibinfo {author} {\bibfnamefont
  {Y.}~\bibnamefont {Andersson}}, \ and\ \bibinfo {author} {\bibfnamefont
  {P.}~\bibnamefont {Nordblad}},\ }\href {\doibase 10.1103/PhysRevB.88.094440}
  {\bibfield  {journal} {\bibinfo  {journal} {Phys. Rev. B}\ }\textbf {\bibinfo
  {volume} {88}},\ \bibinfo {pages} {094440} (\bibinfo {year}
  {2013})}\BibitemShut {NoStop}%
\bibitem [{\citenamefont {van~der Laan}\ and\ \citenamefont
  {Thole}(1991)}]{LAA91}%
  \BibitemOpen
  \bibfield  {author} {\bibinfo {author} {\bibfnamefont {G.}~\bibnamefont
  {van~der Laan}}\ and\ \bibinfo {author} {\bibfnamefont {B.~T.}\ \bibnamefont
  {Thole}},\ }\href {\doibase 10.1103/PhysRevB.43.13401} {\bibfield  {journal}
  {\bibinfo  {journal} {Phys. Rev. B}\ }\textbf {\bibinfo {volume} {43}},\
  \bibinfo {pages} {13401} (\bibinfo {year} {1991})}\BibitemShut {NoStop}%
\bibitem [{\citenamefont {Zaanen}\ \emph {et~al.}(1985)\citenamefont {Zaanen},
  \citenamefont {Sawatzky},\ and\ \citenamefont {Allen}}]{ZAA85}%
  \BibitemOpen
  \bibfield  {author} {\bibinfo {author} {\bibfnamefont {J.}~\bibnamefont
  {Zaanen}}, \bibinfo {author} {\bibfnamefont {G.~A.}\ \bibnamefont
  {Sawatzky}}, \ and\ \bibinfo {author} {\bibfnamefont {J.~W.}\ \bibnamefont
  {Allen}},\ }\href {\doibase 10.1103/PhysRevLett.55.418} {\bibfield  {journal}
  {\bibinfo  {journal} {Phys. Rev. Lett.}\ }\textbf {\bibinfo {volume} {55}},\
  \bibinfo {pages} {418} (\bibinfo {year} {1985})}\BibitemShut {NoStop}%
\bibitem [{\citenamefont {van~der Laan}\ \emph {et~al.}(1986)\citenamefont
  {van~der Laan}, \citenamefont {Zaanen}, \citenamefont {Sawatzky},
  \citenamefont {Karnatak},\ and\ \citenamefont {Esteva}}]{LAA86}%
  \BibitemOpen
  \bibfield  {author} {\bibinfo {author} {\bibfnamefont {G.}~\bibnamefont
  {van~der Laan}}, \bibinfo {author} {\bibfnamefont {J.}~\bibnamefont
  {Zaanen}}, \bibinfo {author} {\bibfnamefont {G.~A.}\ \bibnamefont
  {Sawatzky}}, \bibinfo {author} {\bibfnamefont {R.}~\bibnamefont {Karnatak}},
  \ and\ \bibinfo {author} {\bibfnamefont {J.-M.}\ \bibnamefont {Esteva}},\
  }\href {\doibase 10.1103/PhysRevB.33.4253} {\bibfield  {journal} {\bibinfo
  {journal} {Phys. Rev. B}\ }\textbf {\bibinfo {volume} {33}},\ \bibinfo
  {pages} {4253} (\bibinfo {year} {1986})}\BibitemShut {NoStop}%
\bibitem [{\citenamefont {Jian-Wang}\ \emph {et~al.}(1993)\citenamefont
  {Jian-Wang}, \citenamefont {He-Lie},\ and\ \citenamefont {Qing-Qi}}]{CAI93}%
  \BibitemOpen
  \bibfield  {author} {\bibinfo {author} {\bibfnamefont {C.}~\bibnamefont
  {Jian-Wang}}, \bibinfo {author} {\bibfnamefont {L.}~\bibnamefont {He-Lie}}, \
  and\ \bibinfo {author} {\bibfnamefont {Z.}~\bibnamefont {Qing-Qi}},\ }\href
  {http://stacks.iop.org/0953-8984/5/i=50/a=012} {\bibfield  {journal}
  {\bibinfo  {journal} {J. Phys.: Condens. Matter}\ }\textbf {\bibinfo {volume}
  {5}},\ \bibinfo {pages} {9307} (\bibinfo {year} {1993})}\BibitemShut
  {NoStop}%
\bibitem [{\citenamefont {Thole}\ \emph {et~al.}(1992)\citenamefont {Thole},
  \citenamefont {Carra}, \citenamefont {Sette},\ and\ \citenamefont {van~der
  Laan}}]{THO92}%
  \BibitemOpen
  \bibfield  {author} {\bibinfo {author} {\bibfnamefont {B.~T.}\ \bibnamefont
  {Thole}}, \bibinfo {author} {\bibfnamefont {P.}~\bibnamefont {Carra}},
  \bibinfo {author} {\bibfnamefont {F.}~\bibnamefont {Sette}}, \ and\ \bibinfo
  {author} {\bibfnamefont {G.}~\bibnamefont {van~der Laan}},\ }\href {\doibase
  10.1103/PhysRevLett.68.1943} {\bibfield  {journal} {\bibinfo  {journal}
  {Phys. Rev. Lett.}\ }\textbf {\bibinfo {volume} {68}},\ \bibinfo {pages}
  {1943} (\bibinfo {year} {1992})}\BibitemShut {NoStop}%
\bibitem [{\citenamefont {Carra}\ \emph {et~al.}(1993)\citenamefont {Carra},
  \citenamefont {Thole}, \citenamefont {Altarelli},\ and\ \citenamefont
  {Wang}}]{CAR93}%
  \BibitemOpen
  \bibfield  {author} {\bibinfo {author} {\bibfnamefont {P.}~\bibnamefont
  {Carra}}, \bibinfo {author} {\bibfnamefont {B.~T.}\ \bibnamefont {Thole}},
  \bibinfo {author} {\bibfnamefont {M.}~\bibnamefont {Altarelli}}, \ and\
  \bibinfo {author} {\bibfnamefont {X.}~\bibnamefont {Wang}},\ }\href {\doibase
  10.1103/PhysRevLett.70.694} {\bibfield  {journal} {\bibinfo  {journal} {Phys.
  Rev. Lett.}\ }\textbf {\bibinfo {volume} {70}},\ \bibinfo {pages} {694}
  (\bibinfo {year} {1993})}\BibitemShut {NoStop}%
\bibitem [{\citenamefont {Piamonteze}\ \emph {et~al.}(2009)\citenamefont
  {Piamonteze}, \citenamefont {Miedema},\ and\ \citenamefont
  {de~Groot}}]{PIA09}%
  \BibitemOpen
  \bibfield  {author} {\bibinfo {author} {\bibfnamefont {C.}~\bibnamefont
  {Piamonteze}}, \bibinfo {author} {\bibfnamefont {P.}~\bibnamefont {Miedema}},
  \ and\ \bibinfo {author} {\bibfnamefont {F.~M.~F.}\ \bibnamefont
  {de~Groot}},\ }\href {\doibase 10.1103/PhysRevB.80.184410} {\bibfield
  {journal} {\bibinfo  {journal} {Phys. Rev. B}\ }\textbf {\bibinfo {volume}
  {80}},\ \bibinfo {pages} {184410} (\bibinfo {year} {2009})}\BibitemShut
  {NoStop}%
\bibitem [{\citenamefont {Teramura}\ \emph {et~al.}(1996)\citenamefont
  {Teramura}, \citenamefont {Tanaka},\ and\ \citenamefont {Jo}}]{YOS96}%
  \BibitemOpen
  \bibfield  {author} {\bibinfo {author} {\bibfnamefont {Y.}~\bibnamefont
  {Teramura}}, \bibinfo {author} {\bibfnamefont {A.}~\bibnamefont {Tanaka}}, \
  and\ \bibinfo {author} {\bibfnamefont {T.}~\bibnamefont {Jo}},\ }\href
  {\doibase 10.1143/JPSJ.65.1053} {\bibfield  {journal} {\bibinfo  {journal}
  {J. Phys. Soc. Jpn.}\ }\textbf {\bibinfo {volume} {65}},\ \bibinfo {pages}
  {1053} (\bibinfo {year} {1996})}\BibitemShut {NoStop}%
\bibitem [{\citenamefont {van~der Laan}\ \emph {et~al.}(2004)\citenamefont
  {van~der Laan}, \citenamefont {Moore}, \citenamefont {Tobin}, \citenamefont
  {Chung}, \citenamefont {Wall},\ and\ \citenamefont {Schwartz}}]{LAA04}%
  \BibitemOpen
  \bibfield  {author} {\bibinfo {author} {\bibfnamefont {G.}~\bibnamefont
  {van~der Laan}}, \bibinfo {author} {\bibfnamefont {K.~T.}\ \bibnamefont
  {Moore}}, \bibinfo {author} {\bibfnamefont {J.~G.}\ \bibnamefont {Tobin}},
  \bibinfo {author} {\bibfnamefont {B.~W.}\ \bibnamefont {Chung}}, \bibinfo
  {author} {\bibfnamefont {M.~A.}\ \bibnamefont {Wall}}, \ and\ \bibinfo
  {author} {\bibfnamefont {A.~J.}\ \bibnamefont {Schwartz}},\ }\href {\doibase
  10.1103/PhysRevLett.93.097401} {\bibfield  {journal} {\bibinfo  {journal}
  {Phys. Rev. Lett.}\ }\textbf {\bibinfo {volume} {93}},\ \bibinfo {pages}
  {097401} (\bibinfo {year} {2004})}\BibitemShut {NoStop}%
\bibitem [{\citenamefont {Wu}\ \emph {et~al.}(1993)\citenamefont {Wu},
  \citenamefont {Wang},\ and\ \citenamefont {Freeman}}]{WU93}%
  \BibitemOpen
  \bibfield  {author} {\bibinfo {author} {\bibfnamefont {R.}~\bibnamefont
  {Wu}}, \bibinfo {author} {\bibfnamefont {D.}~\bibnamefont {Wang}}, \ and\
  \bibinfo {author} {\bibfnamefont {A.~J.}\ \bibnamefont {Freeman}},\ }\href
  {\doibase 10.1103/PhysRevLett.71.3581} {\bibfield  {journal} {\bibinfo
  {journal} {Phys. Rev. Lett.}\ }\textbf {\bibinfo {volume} {71}},\ \bibinfo
  {pages} {3581} (\bibinfo {year} {1993})}\BibitemShut {NoStop}%
\bibitem [{\citenamefont {Crocombette}\ \emph {et~al.}(1996)\citenamefont
  {Crocombette}, \citenamefont {Thole},\ and\ \citenamefont {Jollet}}]{CRO96}%
  \BibitemOpen
  \bibfield  {author} {\bibinfo {author} {\bibfnamefont {J.~P.}\ \bibnamefont
  {Crocombette}}, \bibinfo {author} {\bibfnamefont {B.~T.}\ \bibnamefont
  {Thole}}, \ and\ \bibinfo {author} {\bibfnamefont {F.}~\bibnamefont
  {Jollet}},\ }\href {http://stacks.iop.org/0953-8984/8/i=22/a=013} {\bibfield
  {journal} {\bibinfo  {journal} {J. Phys.: Condens. Matter}\ }\textbf
  {\bibinfo {volume} {8}},\ \bibinfo {pages} {4095} (\bibinfo {year}
  {1996})}\BibitemShut {NoStop}%
\bibitem [{\citenamefont {Chen}\ \emph {et~al.}(1995)\citenamefont {Chen},
  \citenamefont {Idzerda}, \citenamefont {Lin}, \citenamefont {Smith},
  \citenamefont {Meigs}, \citenamefont {Chaban}, \citenamefont {Ho},
  \citenamefont {Pellegrin},\ and\ \citenamefont {Sette}}]{CHEN95}%
  \BibitemOpen
  \bibfield  {author} {\bibinfo {author} {\bibfnamefont {C.~T.}\ \bibnamefont
  {Chen}}, \bibinfo {author} {\bibfnamefont {Y.~U.}\ \bibnamefont {Idzerda}},
  \bibinfo {author} {\bibfnamefont {H.-J.}\ \bibnamefont {Lin}}, \bibinfo
  {author} {\bibfnamefont {N.~V.}\ \bibnamefont {Smith}}, \bibinfo {author}
  {\bibfnamefont {G.}~\bibnamefont {Meigs}}, \bibinfo {author} {\bibfnamefont
  {E.}~\bibnamefont {Chaban}}, \bibinfo {author} {\bibfnamefont {G.~H.}\
  \bibnamefont {Ho}}, \bibinfo {author} {\bibfnamefont {E.}~\bibnamefont
  {Pellegrin}}, \ and\ \bibinfo {author} {\bibfnamefont {F.}~\bibnamefont
  {Sette}},\ }\href {\doibase 10.1103/PhysRevLett.75.152} {\bibfield  {journal}
  {\bibinfo  {journal} {Phys. Rev. Lett.}\ }\textbf {\bibinfo {volume} {75}},\
  \bibinfo {pages} {152} (\bibinfo {year} {1995})}\BibitemShut {NoStop}%
\bibitem [{\citenamefont {van~der Laan}\ \emph {et~al.}(2010)\citenamefont
  {van~der Laan}, \citenamefont {Edmonds}, \citenamefont {Arenholz},
  \citenamefont {Farley},\ and\ \citenamefont {Gallagher}}]{LAA10}%
  \BibitemOpen
  \bibfield  {author} {\bibinfo {author} {\bibfnamefont {G.}~\bibnamefont
  {van~der Laan}}, \bibinfo {author} {\bibfnamefont {K.~W.}\ \bibnamefont
  {Edmonds}}, \bibinfo {author} {\bibfnamefont {E.}~\bibnamefont {Arenholz}},
  \bibinfo {author} {\bibfnamefont {N.~R.~S.}\ \bibnamefont {Farley}}, \ and\
  \bibinfo {author} {\bibfnamefont {B.~L.}\ \bibnamefont {Gallagher}},\ }\href
  {\doibase 10.1103/PhysRevB.81.214422} {\bibfield  {journal} {\bibinfo
  {journal} {Phys. Rev. B}\ }\textbf {\bibinfo {volume} {81}},\ \bibinfo
  {pages} {214422} (\bibinfo {year} {2010})}\BibitemShut {NoStop}%
\bibitem [{\citenamefont {Kronast}\ \emph {et~al.}(2006)\citenamefont
  {Kronast}, \citenamefont {Ovsyannikov}, \citenamefont {Vollmer},
  \citenamefont {D\"urr}, \citenamefont {Eberhardt}, \citenamefont {Imperia},
  \citenamefont {Schmitz}, \citenamefont {Schott}, \citenamefont {Ruester},
  \citenamefont {Gould}, \citenamefont {Schmidt}, \citenamefont {Brunner},
  \citenamefont {Sawicki},\ and\ \citenamefont {Molenkamp}}]{KRO06}%
  \BibitemOpen
  \bibfield  {author} {\bibinfo {author} {\bibfnamefont {F.}~\bibnamefont
  {Kronast}}, \bibinfo {author} {\bibfnamefont {R.}~\bibnamefont
  {Ovsyannikov}}, \bibinfo {author} {\bibfnamefont {A.}~\bibnamefont
  {Vollmer}}, \bibinfo {author} {\bibfnamefont {H.~A.}\ \bibnamefont {D\"urr}},
  \bibinfo {author} {\bibfnamefont {W.}~\bibnamefont {Eberhardt}}, \bibinfo
  {author} {\bibfnamefont {P.}~\bibnamefont {Imperia}}, \bibinfo {author}
  {\bibfnamefont {D.}~\bibnamefont {Schmitz}}, \bibinfo {author} {\bibfnamefont
  {G.~M.}\ \bibnamefont {Schott}}, \bibinfo {author} {\bibfnamefont
  {C.}~\bibnamefont {Ruester}}, \bibinfo {author} {\bibfnamefont
  {C.}~\bibnamefont {Gould}}, \bibinfo {author} {\bibfnamefont
  {G.}~\bibnamefont {Schmidt}}, \bibinfo {author} {\bibfnamefont
  {K.}~\bibnamefont {Brunner}}, \bibinfo {author} {\bibfnamefont
  {M.}~\bibnamefont {Sawicki}}, \ and\ \bibinfo {author} {\bibfnamefont
  {L.~W.}\ \bibnamefont {Molenkamp}},\ }\href {\doibase
  10.1103/PhysRevB.74.235213} {\bibfield  {journal} {\bibinfo  {journal} {Phys.
  Rev. B}\ }\textbf {\bibinfo {volume} {74}},\ \bibinfo {pages} {235213}
  (\bibinfo {year} {2006})}\BibitemShut {NoStop}%
\bibitem [{\citenamefont {Liu}\ and\ \citenamefont
  {Altounian}(2009)}]{XBLIU09}%
  \BibitemOpen
  \bibfield  {author} {\bibinfo {author} {\bibfnamefont {X.~B.}\ \bibnamefont
  {Liu}}\ and\ \bibinfo {author} {\bibfnamefont {Z.}~\bibnamefont
  {Altounian}},\ }\href {\doibase http://dx.doi.org/10.1063/1.3056408}
  {\bibfield  {journal} {\bibinfo  {journal} {J. Appl. Phys.}\ }\textbf
  {\bibinfo {volume} {105}},\ \bibinfo {eid} {07A902} (\bibinfo {year}
  {2009})}\BibitemShut {NoStop}%
\bibitem [{\citenamefont {Hudl}\ \emph {et~al.}(2011)\citenamefont {Hudl},
  \citenamefont {Nordblad}, \citenamefont {Bj\"orkman}, \citenamefont
  {Eriksson}, \citenamefont {H\"aggstr\"om}, \citenamefont {Sahlberg},
  \citenamefont {Andersson}, \citenamefont {Delczeg-Czirjak},\ and\
  \citenamefont {Vitos}}]{HUD11}%
  \BibitemOpen
  \bibfield  {author} {\bibinfo {author} {\bibfnamefont {M.}~\bibnamefont
  {Hudl}}, \bibinfo {author} {\bibfnamefont {P.}~\bibnamefont {Nordblad}},
  \bibinfo {author} {\bibfnamefont {T.}~\bibnamefont {Bj\"orkman}}, \bibinfo
  {author} {\bibfnamefont {O.}~\bibnamefont {Eriksson}}, \bibinfo {author}
  {\bibfnamefont {L.}~\bibnamefont {H\"aggstr\"om}}, \bibinfo {author}
  {\bibfnamefont {M.}~\bibnamefont {Sahlberg}}, \bibinfo {author}
  {\bibfnamefont {Y.}~\bibnamefont {Andersson}}, \bibinfo {author}
  {\bibfnamefont {E.-K.}\ \bibnamefont {Delczeg-Czirjak}}, \ and\ \bibinfo
  {author} {\bibfnamefont {L.}~\bibnamefont {Vitos}},\ }\href {\doibase
  10.1103/PhysRevB.83.134420} {\bibfield  {journal} {\bibinfo  {journal} {Phys.
  Rev. B}\ }\textbf {\bibinfo {volume} {83}},\ \bibinfo {pages} {134420}
  (\bibinfo {year} {2011})}\BibitemShut {NoStop}%
\bibitem [{\citenamefont {Guillou}\ \emph
  {et~al.}(2014{\natexlab{b}})\citenamefont {Guillou}, \citenamefont {Yibole},
  \citenamefont {Porcari}, \citenamefont {Zhang}, \citenamefont {van Dijk},\
  and\ \citenamefont {Br\"uck}}]{GUI14B}%
  \BibitemOpen
  \bibfield  {author} {\bibinfo {author} {\bibfnamefont {F.}~\bibnamefont
  {Guillou}}, \bibinfo {author} {\bibfnamefont {H.}~\bibnamefont {Yibole}},
  \bibinfo {author} {\bibfnamefont {G.}~\bibnamefont {Porcari}}, \bibinfo
  {author} {\bibfnamefont {L.}~\bibnamefont {Zhang}}, \bibinfo {author}
  {\bibfnamefont {N.~H.}\ \bibnamefont {van Dijk}}, \ and\ \bibinfo {author}
  {\bibfnamefont {E.}~\bibnamefont {Br\"uck}},\ }\href {\doibase
  http://dx.doi.org/10.1063/1.4892406} {\bibfield  {journal} {\bibinfo
  {journal} {J. Appl. Phys.}\ }\textbf {\bibinfo {volume} {116}},\ \bibinfo
  {eid} {063903} (\bibinfo {year} {2014}{\natexlab{b}})}\BibitemShut {NoStop}%
\bibitem [{\citenamefont {Fujii}\ \emph {et~al.}(1988)\citenamefont {Fujii},
  \citenamefont {Uwatoko}, \citenamefont {Motoya}, \citenamefont {Ito},\ and\
  \citenamefont {Okamoto}}]{FUJ88}%
  \BibitemOpen
  \bibfield  {author} {\bibinfo {author} {\bibfnamefont {H.}~\bibnamefont
  {Fujii}}, \bibinfo {author} {\bibfnamefont {Y.}~\bibnamefont {Uwatoko}},
  \bibinfo {author} {\bibfnamefont {K.}~\bibnamefont {Motoya}}, \bibinfo
  {author} {\bibfnamefont {Y.}~\bibnamefont {Ito}}, \ and\ \bibinfo {author}
  {\bibfnamefont {T.}~\bibnamefont {Okamoto}},\ }\href {\doibase
  10.1143/JPSJ.57.2143} {\bibfield  {journal} {\bibinfo  {journal} {J. Phys.
  Soc. Jpn.}\ }\textbf {\bibinfo {volume} {57}},\ \bibinfo {pages} {2143}
  (\bibinfo {year} {1988})}\BibitemShut {NoStop}%
\bibitem [{\citenamefont {Zach}\ \emph {et~al.}(1990)\citenamefont {Zach},
  \citenamefont {Guillot},\ and\ \citenamefont {Fruchart}}]{ZAC90}%
  \BibitemOpen
  \bibfield  {author} {\bibinfo {author} {\bibfnamefont {R.}~\bibnamefont
  {Zach}}, \bibinfo {author} {\bibfnamefont {M.}~\bibnamefont {Guillot}}, \
  and\ \bibinfo {author} {\bibfnamefont {R.}~\bibnamefont {Fruchart}},\ }\href
  {\doibase http://dx.doi.org/10.1016/0304-8853(90)90730-E} {\bibfield
  {journal} {\bibinfo  {journal} {J. Magn. Magn. Mater.}\ }\textbf {\bibinfo
  {volume} {89}},\ \bibinfo {pages} {221 } (\bibinfo {year}
  {1990})}\BibitemShut {NoStop}%
\end{thebibliography}

%merlin.mbs apsrev4-1.bst 2010-07-25 4.21a (PWD, AO, DPC) hacked
%Control: key (0)
%Control: author (8) initials jnrlst
%Control: editor formatted (1) identically to author
%Control: production of article title (-1) disabled
%Control: page (0) single
%Control: year (1) truncated
%Control: production of eprint (0) enabled
%merlin.mbs apsrev4-1.bst 2010-07-25 4.21a (PWD, AO, DPC) hacked
%Control: key (0)
%Control: author (8) initials jnrlst
%Control: editor formatted (1) identically to author
%Control: production of article title (-1) disabled
%Control: page (0) single
%Control: year (1) truncated
%Control: production of eprint (0) enabled
\providecommand{\noopsort}[1]{}\providecommand{\singleletter}[1]{#1}%

\end{document}